\documentclass[prd,a4paper,twocolumn,preprintnumbers,nofootinbib]{revtex4}

\usepackage{amsmath,amssymb,slashed,mathtools,slashed}
\usepackage{graphicx,multirow}
\usepackage{units}
\usepackage[hyperfootnotes=false,colorlinks,citecolor=blue]{hyperref}
\usepackage{mathtools,halloweenmath}
\usepackage{pifont}

\newcommand{\beq}{\begin{equation}}
\newcommand{\eeq}{\end{equation}}
\newcommand{\bea}{\begin{eqnarray}}
\newcommand{\ena}{\end{eqnarray}}

\def \epsilon {\varepsilon} 

\allowdisplaybreaks

\begin{document}

\title{Constraining Lepton Flavor Violating Higgs Couplings at the HL-LHC in the Vector Boson Fusion Channel
}

\author{Rahool Kumar Barman}
\email{rahool.barman@okstate.edu}
\affiliation{Department of Physics, Oklahoma State University, Stillwater, OK 74078, USA}

\author{P. S. Bhupal Dev}
\email{bdev@wustl.edu}
\affiliation{Department of Physics and McDonnell Center for the Space Sciences, Washington University, St. Louis, MO 63130, USA}

\author{ Anil Thapa}
\email{wtd8kz@virginia.edu}
\affiliation{Department of Physics, University of Virginia, Charlottesville, VA 22904-4714, USA}

\begin{abstract}
We explore the parameter space of lepton flavor violating (LFV) neutral Higgs Yukawa couplings with the muon and tau leptons that can be probed at the high-luminosity Large Hadron Collider (HL-LHC) via the vector boson fusion~(VBF) Higgs production process. Our projected sensitivities for the Standard Model Higgs ($h$) LFV branching ratio ${\rm Br}(h \to \mu\tau)$ in the $pp \to h j j \to (h \to \mu \tau) jj$ channel at the HL-LHC are contrasted with the current and future low-energy constraints from the anomalous magnetic moment and electric dipole moment of the muon, as well as with other LFV observables, such as $\tau\to 3\mu$ and $\tau\to \mu\gamma$. We also study the LFV prospects of a generic beyond the Standard Model neutral Higgs boson ($H$) with a mass in the range of $m_{H}\in [20,800]~$GeV and give the projected model-independent upper limits on the VBF production cross-section of $Hjj$ times the branching ratio of $H\to \mu\tau$ at the HL-LHC. We interpret these results in the context of a two-Higgs doublet model as a case study. 

\end{abstract}

\maketitle

\section{Introduction}

The discovery of the Higgs boson of mass 125 GeV at the LHC~\cite{ATLAS:2012yve,CMS:2012qbp} has opened the possibility to gain deeper insight into the mechanism of electroweak symmetry breaking (EWSB) and to search for beyond the Standard Model (BSM) physics phenomena in the Higgs sector. Although the properties of the 125 GeV Higgs boson are thus far consistent with the Standard Model (SM) expectations~\cite{ATLAS:2022vkf, CMS:2022dwd}, any statistically significant deviations from the SM predictions in future data could point to a new source of EWSB or some BSM physics close to the electroweak scale. Therefore, it is important to search for nonstandard processes involving the Higgs boson. 

Within the SM, the Higgs boson couplings to fermions are flavor diagonal:  $Y_{ij} = (m_i/v) \delta_{ij}$, where $v=246$ GeV is the electroweak vacuum expectation value and $m_i$ are the fermion masses. However, these couplings can be quite different in the presence of new physics. In particular, there exist several BSM scenarios which allow for lepton flavor violating (LFV) couplings of the Higgs boson which are absent in the SM; see e.g.  Refs.~\cite{Bjorken:1977vt, Diaz-Cruz:1999sns,Raidal:2008jk,Blankenburg:2012ex, Harnik:2012pb,deGouvea:2013zba,Vicente:2015cka,Lindner:2016bgg}.

In the case when the 125 GeV Higgs boson is the only source for EWSB and the BSM physics present in the model consists of heavy fields that can be integrated out~\cite{Buchmuller:1985jz,Babu:1999me,Giudice:2008uua,Agashe:2009di}, the Yukawa coupling in the mass basis after EWSB can be written as
\begin{equation}
    Y_{ij} = \frac{m_i}{v} \delta_{ij} + \frac{v^2}{\sqrt{2} \Lambda^2} \lambda_{ij} \, ,
    \label{eq:Yuk}
\end{equation}
where $\Lambda$ is the scale of new physics and $\lambda_{ij}$ are the coefficients associated with the lowest-dimension (dimension-6) effective operators that modify the Yukawa interactions, namely~\cite{Herrero-Garcia:2016uab}
\begin{align}
{\cal L} \supset -\frac{\lambda_{ij}}{\Lambda^2} (\bar{f_L}^i f_R^j) \phi (\phi^\dagger \phi)+{\rm H.c.}, 
\end{align}
where $\phi$ is the SM Higgs doublet field and $f_L,f_R$ are the left- and right-handed SM fermions respectively. Here $\lambda$ is in principle an arbitrary non-diagonal matrix that can significantly modify the Higgs Yukawa interactions for $\Lambda$ of the order of electroweak scale. It is worth pointing out that to reproduce the hierarchical spectrum of the SM fermion masses, we need to impose either fine-tuning in Eq.~\eqref{eq:Yuk} or the naturalness condition for the off-diagonal couplings~\cite{Cheng:1987rs} 
\begin{align}
|Y_{ij} Y_{ji}| \lesssim \frac{m_i m_j}{v^2} \quad {\rm for}~i \neq j. 
\label{eq:natural}
\end{align}

In the case of an additional Higgs boson $\phi_2$ taking part in EWSB, the $\phi_2$ boson with the same quantum numbers as $\phi :({\bf 2},1/2)$ under $SU(2)_L \times U(1)_Y$ can contribute to the quark and lepton masses. This allows the Yukawa couplings of the 125 GeV Higgs boson to be misaligned with respect to the SM. In addition, the new scalar if leptophilic can remain sufficiently light and lead to sizable LFV while satisfying the constraints from flavor changing neutral current (FCNC) as well collider bounds. A concrete example is the lepton-specific two Higgs doublet model (2HDM)~\cite{Lee:1973iz,Branco:2011iw,Crivellin:2015hha}.   

Without loss of generality and referring to any specific model, we write the effective LFV Yukawa couplings of the (B)SM Higgs boson to the charged leptons as
\begin{equation}
    \mathcal{L}_Y \supset -Y_{ij}\ \bar{\ell}_{Li} e_{Rj} h~(H) + {\rm H.c.} 
    \label{eq:lagY}
\end{equation}
with $Y_{ij}$ ($i\neq j$) as free parameters. 
We set the diagonal couplings $Y_{ii}$ to their respective SM values, i.e.~$Y_{ii} = m_i/v$. The new interactions in Eq.~\eqref{eq:lagY} can lead to new LFV Higgs decay modes that may be directly observable in current and future collider experiments. The ATLAS and CMS collaborations have performed several searches to study the LFV decays of the SM Higgs boson in the $h \to e\mu$~\cite{CMS:2016cvq}, $e\tau$~\cite{ATLAS:2016joj,CMS:2016cvq, CMS:2017con,ATLAS:2019pmk,CMS:2021rsq} and $\mu\tau$~\cite{CMS:2015qee, ATLAS:2015cji,ATLAS:2016joj, CMS:2017con, ATLAS:2019pmk,CMS:2021rsq} channels at the LHC; however, any significant excess over SM expectations is yet to be observed. LFV decays of neutral heavy resonances and heavy Higgs bosons at the LHC have also been investigated in Refs.~\cite{ATLAS:2015dva, ATLAS:2016loq, CMS:2018hnz, ATLAS:2018mrn, CMS:2022fsw} and \cite{LHCb:2018ukt,CMS:2019pex}, respectively. The prospects of probing LFV signals induced by Higgs at the future lepton~\cite{Chakraborty:2016gff,Dev:2017ftk,Chakraborty:2017tyb,Qin:2017aju,Li:2018cod} and hadron~\cite{Bhattacherjee:2015sia, Banerjee:2016foh,Huitu:2016pwk,Arganda:2019gnv,Chattopadhyay:2019ycs,Ghosh:2020tfq,Ghosh:2021jeg, Asai:2021ugj, Arroyo-Urena:2020mgg, Hou:2022nyh} colliders have also been widely explored. 

In this work, we focus on the LFV Higgs signal in the $\mu\tau$ channel that can be effectively probed in a model-independent way at the HL-LHC using the vector boson fusion (VBF) channel. We first study the projected sensitivity for LFV decays of the SM-like Higgs boson, $h \to \mu\tau$, in the VBF Higgs production channel, $pp \to (h \to \mu\tau) jj$, at the HL-LHC~($\sqrt{s}=14~$TeV, $\mathcal{L}=3~{\rm ab}^{-1}$). We then perform a detailed collider analysis to explore LFV decays of a BSM Higgs boson $H$ in the VBF production channel, $pp \to (H \to \mu\tau)jj $ for several BSM Higgs masses $m_{H}$, and derive `model-agnostic' projected upper limits on the production cross-section times ${\rm Br}(H \to \mu\tau)$ for $m_{H} \in [20,800]~$GeV.

Typically, at the hadron colliders, the major impetus has been on gluon gluon fusion~(ggF) Higgs production mode, while LFV Higgs decays in the VBF channel have been much less explored. This bias is understandable since the ggF production rates are much larger than VBF in a typical SM Higgs-like scenario with $m_{h} \sim 125~$GeV. 

As expected, the leading sensitivity in searches for LFV decays of $h$ arises from the non-VBF Higgs production category, largely constituted by ggF production. This is also highlighted in the recent searches by CMS~\cite{CMS:2021rsq} and ATLAS~\cite{ATLAS:2019pmk}, where the limits from VBF signal regions alone are weaker by a factor of few than their non-VBF counterparts. However, the VBF production channel becomes extremely relevant in new physics scenarios with extended Higgs sectors like the singlet and 2HDM extensions. The ggF and VBF production rates become comparable for heavier BSM Higgs states $H$ at masses closer to $\mathcal{O}(1)~$TeV~\cite{ LHCHiggsCrossSectionWorkingGroup:2011wcg, Das:2018fog}. In addition, the distinct phenomenological features of the VBF topology offer better control for signal-to-background discrimination than the ggF signal. Overall, the VBF channel can play a complementary role, if not leading, in the search for LFV decays of BSM Higgs extensions and may lead to exciting theoretical implications for new physics, as we show in this work. 

It is worth noting that LFV decays of the (B)SM Higgs boson can also be realized in other Higgs production channels, such as Higgstrahlung process $pp \to Zh(H)$. For a SM-like Higgs scenario with $m_{h} \sim 125~$GeV, the VBF production cross-section is 4.4 times larger than the $Zh$ production rate at the $\sqrt{s}=14~$TeV LHC~\cite{LHCHiggsCrossSectionWorkingGroup:2011wcg}. The disparity grows wider at higher Higgs masses; for instance, at $m_{H} \sim 1~$TeV, the VBF to $ZH$ cross-section ratio is $\sim 292$~\cite{LHCHiggsCrossSectionWorkingGroup:2011wcg}. Because of considerably smaller cross-sections, especially at heavier Higgs masses, the sensitivity for LFV decays of BSM Higgs bosons in the Higgstrahlung channel is expected to be sub-leading than in ggF and VBF modes in a typical 2HDM extension. Therefore, the $Zh$ mode is not considered in the present analysis.

As for the LFV signal itself, ideally, all three LFV decay channels, $h/H \to e\mu, e\tau$ and $\mu\tau$, should be considered in the search for LFV decays of the (B)SM Higgs bosons. However, the partial decay width of an SM Higgs-like boson into the LFV final states is typically proportional to the mass of the heavier lepton, resulting in a usually suppressed signal rate for the $e\mu$ channel than $e\tau$ and $\mu\tau$. Additionally, the rare $\mu$ decay processes typically impose stringent upper limits on $Y_{e\mu}$, thus further restricting the search potential in the $e\mu$ channel. The $e\tau$ and $\mu\tau$ decay channels result in roughly similar sensitivity at the LHC~\cite{CMS:2021rsq}; however, the background simulation for the $e\tau$ channel is more challenging due to relatively more significant contamination from the non-prompt-lepton backgrounds. Moreover, we would also like to connect the LFV signal at HL-LHC with the precision low-energy observable of muon anomalous magnetic moment~\cite{Muong-2:2021ojo}, for which the loop contribution involving a $\mu\tau$ flavor-violating Higgs is typically enhanced by a factor of $m_\tau/m_\mu$~\cite{Lindner:2016bgg}. In addition, we find that the low-energy constraint from muon electric dipole moment (EDM) on $Y_{\mu\tau}$ is 10 orders of magnitude less stringent than that on $Y_{e\tau}$ from electron EDM~\cite{ACME:2018yjb}, thus making the collider study of the $\mu\tau$ channel more relevant. Due to the above reasons, we only focus on LFV decays of the Higgs boson in the $\mu\tau$ channel.

The rest of the paper is organized as follows. In Sec.~\ref{sec:lowenergy} we present various low energy LFV constraints on the Yukawa couplings $Y_{\mu\tau}$ and $Y_{\tau\mu}$. Sec.~\ref{sec:proj} discusses the projected sensitivity of the LFV couplings of the SM Higgs boson at the HL-LHC in the VBF channel. Sec.~\ref{sec:BSM} discusses the HL-LHC reach for a generic BSM Higgs LFV decay, as well as a specific example in 2HDM. We conclude in Sec.~\ref{sec:conclusion}.  

\section{Low-energy Constraints}\label{sec:lowenergy}

The LFV couplings in Eq.~\eqref{eq:lagY} are subject to various low energy constraints discussed below.

\subsection{Dipole Moment}
The CP-violating and conserving parts of the Yukawa couplings lead to the electric and magnetic dipole moment of the leptons. The flavor violating neutral Higgs contribution to the anomalous magnetic moment $(g-2)_\mu$ at one loop~\cite{Leveille:1977rc} in the limit $m_i < m_H$ is given by 
\begin{equation}
    \Delta a_\mu \simeq \frac{\Re(Y_{\mu i}\ Y_{i \mu})}{4 \pi^2} \frac{m_\mu m_i}{m_H^2} \left( -\frac{3}{4} + \log \frac{m_H}{m_i}\right) \, .
    \label{eq:g2}
\end{equation}
The difference between the experimentally measured value~\cite{Muong-2:2006rrc, Muong-2:2021ojo} and the theoretical one predicted by the SM~\cite{Aoyama:2020ynm}, $\Delta a_\mu = a_\mu^{\rm exp} - a_\mu^{\rm SM} = (251 \pm 59) \times 10^{-11}$, is of $4.2\sigma$ discrepancy.\footnote{It should be noted here that a recent lattice simulation result from the BMW collaboration~\cite{Borsanyi:2020mff} is more consistent with the experimental value~\cite{Muong-2:2021ojo}. Moreover, recent results from other lattice groups seem to be converging towards the BMW result~\cite{Ce:2022kxy,Alexandrou:2022amy}. However, these results are in tension with the low-energy $\sigma(e^+e^-\to{\rm hadrons})$ data~\cite{Crivellin:2020zul,Keshavarzi:2020bfy,Colangelo:2020lcg}, and further investigations are ongoing. Until the dust is settled, we choose to use the discrepancy quoted in Ref.~\cite{Muong-2:2021ojo}.} In our case, the dominant contribution arises from a $\tau$-Higgs loop and leads to the relation: $\Re(Y_{\mu\tau}\ Y_{\tau \mu}) \simeq (2.37 \pm 0.56) \times 10^{-3}$ in order to accommodate the $(g-2)_\mu$ anomaly at $1\sigma$ for $m_H=125$ GeV. 

On the other hand, the EDM of leptons places constraints on the imaginary part of the Yukawa couplings of the Higgs field. These constraints are only significant when there is a chirality flip in the fermion line inside the loop. Neglecting terms suppressed by $m_\mu/m_\tau$ and $m_\ell/m_H$, muon EDM is given by~\cite{Ecker:1983dj} 
\begin{equation}
d_{\mu} \simeq-\frac{\Im\left(Y_{\mu \tau} Y_{\tau \mu}\right)}{4 \pi^{2}} \frac{e\ m_{\tau}}{2 m_H^{2}}\left(-\frac{3}{4}+\log \frac{m_H}{{m_\tau}}\right) \, . 
\end{equation}
The current upper limit from $\mu$EDM measurements $d_{\mu} \leq 1.9 \times 10^{-19}$ e-cm~\cite{Muong-2:2008ebm} translates to an upper bound of $\Im (Y_{\mu\tau} Y_{\tau \mu}) < 1.9$ for $m_H=125$ GeV. The sensitivity reach of the future projection of $\mu$EDM is of the order of  $10^{-22}$ e-cm~\cite{Abe:2019thb,Sato:2021aor,Adelmann:2021udj}, corresponding to $\Im (Y_{\mu\tau} Y_{\tau \mu}) < 6 \times  10^{-4}$. The current limits on tau EDM~\cite{ARGUS:2000riz,Belle:2002nla} and future projections~\cite{Bernreuther:2021elu} are a couple of orders of magnitude weaker than those for muons.  

\subsection{$\ell_i \to \ell_j \gamma$}
It is important to point out that the off-diagonal Yukawa couplings $Y_{ij}$ suffer strong constraints from radiative decays like $\tau \to \mu \gamma$. The general expression for the rate of $\ell_1 \to \ell_2 \gamma$ decay involving the neutral Higgs and a lepton $\ell$ in the loop reads as~\cite{Lavoura:2003xp}
\begin{equation}
\begin{aligned}
    \Gamma_{\ell_1 \to \ell_2 \gamma}^{1{\rm- loop}}=&\frac{\alpha_{\rm em}}{144\,(16\pi^2)^2}\frac{m_1^5}{16 m_{H}^4}\Big[\big(\left|Y_{2\ell}Y_{1\ell}^*\right|^2+\left|Y_{\ell1}Y_{\ell2}^*\right|^2\big)\mathcal{F}^2_1(t)\\
    &\qquad \qquad +\frac{9m^2_\ell}{m^2_1}\big(\left|Y_{1\ell}^*Y_{\ell2}^*\right|^2+\left|Y_{2\ell}Y_{\ell1}\right|^2\big)\mathcal{F}^2_2(t)\Big],
    \end{aligned}\label{radiative1}
\end{equation}
where $t=m_\ell^2/m_H^2$, and 
\begin{equation}
    \begin{aligned}
    \mathcal{F}_1(t)=&\frac{2+3t-6t^2+t^3+6t\log{t}}{(t-1)^4},\\
    \mathcal{F}_2(t)=&\frac{3-4t+t^2+2\log{t}}{(t-1)^3}.
    \end{aligned}\label{radiative2}
\end{equation}
The second term in Eq.~\eqref{radiative1} appears from the chirally enhanced radiative diagrams, whereas the first term has no chirality flip in the fermion line inside the loop. The bounds on the Yukawa couplings as a function of the mediator masses can be derived from the current bound on ${\rm Br}(\tau\to \mu\gamma)<4.4\times 10^{-8}$~\cite{BaBar:2009hkt}. The dominant contribution for one loop arises from a chirally enhanced $\tau$-Higgs loop, giving rise to the constraint $\sqrt{|Y_{\mu\tau}|^2+|Y_{\tau\mu}|^2} < 0.17$ for $m_H=125$ GeV. 

In addition to the one loop contribution to the LFV process $\tau \to \mu \gamma$, the two loop Barr-Zee diagrams are also significant, where the dominant contribution arises from top-Higgs and $W$-Higgs loops~\cite{Harnik:2012pb}. The relevant rate for $\tau \to \mu \gamma$ reads as 
\begin{equation}
    \Gamma_{\tau \to \mu \gamma}^{\rm 2-{\rm loop}} = \frac{\alpha m_\tau^5}{64 \pi^4}  \frac{|Y_{\tau\mu}^*|^2 + |Y_{\mu\tau}|^2}{m_H^4} (-0.082\ Y_{t} + 0.11)^2 \, ,
    \label{eq:LiLj2loop}
\end{equation}
where $Y_{t} = m_{\rm top}/v$ is the top-quark Yukawa coupling for $m_H=125$ GeV being the SM Higgs mass. 
In Eq.~\eqref{eq:LiLj2loop}, the two terms with relative minus sign respectively represent the terms with top quark and $W$ boson contribution. From this equation, we get $\sqrt{|Y_{\mu\tau}|^2+|Y_{\tau\mu}|^2} < 1.97 \times 10^{-2}$. The full expression can be found in Ref.~\cite{Harnik:2012pb}.

\subsection{Trilepton decay}
In addition to the radiative decays, the flavor changing Higgs boson allows tree level trilepton decay $l_{i}\to l_{k} \bar{l}_{j} l_{l}$. In the limit of massless decay products, the partial decay rate reads as~\cite{Cai:2017jrq}:
\begin{align}
    \Gamma_{l_{i}\to l_{k} \bar{l}_{j} l_{l}} &=\frac{1}{6144 \pi^3}\frac{m_i^5}{4m_H^4}\Big(\frac{1}{(1+\delta_{lk})}(\left|Y^*_{ik}Y^*_{jl}\right|^2+\left|Y_{ki}Y_{lj}\right|^2)\nonumber\\
    &~~~~~~~~~~~~~~~~~~~~~~~+\left|Y^*_{ik}Y_{lj}\right|^2+\left|Y_{ki}Y^*_{jl}\right|^2\Big).
\end{align}
Here $\delta_{lk}$ is the symmetry factor. Using the total tau decay width $\Gamma_\tau^{\rm tot} = 2.27 \times 10^{-12}$ GeV and muon Yukawa coupling $Y_{\mu\mu} = m_\mu/v$, we obtain a bound of  $\sqrt{|Y_{\mu\tau}|^2 + |Y_{\tau\mu}|^2} < 1.35$ for the SM Higgs case from the experimental limit ${\rm Br}(\tau\to 3\mu)<2.1\times 10^{-8}$~\cite{Hayasaka:2010np}. It is clear that these tree level decays are suppressed by the muon Yukawa coupling. On the other hand, loop level contributions do not have such suppression and can be dominant. One loop contribution for $\tau\to 3\mu$ is obtained by attaching a muon line to the photon in the radiative decay of $\tau \to \mu \gamma$, which corresponds to a bound of $\sqrt{|Y_{\mu\tau}|^2 + |Y_{\tau\mu}|^2} < 0.13$~\cite{Goto:2015iha, Harnik:2012pb}. This however turns out to be weaker than the $\tau \to \mu \gamma$ constraint, as expected. 

\subsection{$Z$-boson decay}
In the presence of the Yukawa couplings $Y_{\tau\mu}$ and $Y_{\mu\tau}$, the effective $\mu-\tau-Z$ vertices are induced at one loop order~\cite{Dam:2018rfz}:
\begin{align}
    \Gamma_{Z\to \tau\mu} &= \frac{m_Z}{6\pi}\ ( \frac{1}{2}  |C_L^Z(m_Z^2)|^2 + \frac{m_Z^2}{m_\tau^2} |D_L^Z(m_Z^2)|^2 \nonumber \\
    &~~~+ (L \leftrightarrow R ) ) \, ,
\end{align}
where the coefficients read as
\begin{align}
    C_L^Z(s) &= \frac{g Y_{\tau\tau} Y_{\tau\mu}}{64 \pi^2} (F_V^v (s) g_V^e + F_V^a (s) g_A^e), \nonumber \\
    D_L^Z (s) &= \frac{g Y_{\tau\tau} Y_{\mu\tau}^*}{64 \pi^2} (F_D^v (s) g_V^e + F_D^a (s) g_A^e) .
\end{align}
$C_R^Z$ and $D_R^Z$ are obtained by interchanging $Y_{\tau\mu} \leftrightarrow Y_{\mu\tau}^*$ and $F_{V,D}^a\to - F_{V,D}^a$. The functions $F_{V,D}^{v,a}$ are expressed in terms of Passarino-Veltman functions (see Ref.~\cite{Dam:2018rfz} for details). Numerically the functions $\{F_V^a,F_V^v,F_D^a,F_D^v \}$ at $s=m_Z^2$ read as $\{5-0.78 i, -4.8-0.78i, (-8.1+1.6 i)\times 10^{-5}, 0.84\}$. Using these values and current experimental bound ${\rm Br} (Z \to \mu \tau) < 9.5 \times 10^{-6}$~\cite{ ATLAS:2020zlz} leads to the relation: ${\rm Br}(Z \to \mu^\mp \tau^\pm) = 8.9 \times 10^{-10} |Y_{\mu\tau}|^2 + 7.7 \times 10^{-10} |Y_{\tau\mu}|^2$~\cite{Dam:2018rfz,Goto:2015iha}. The upcoming $e^+e^-$ collider such as the FCC-ee has the sensitivity that can prove the LFV decay of Z in $\mu\tau$ decay up to $\mathcal{O} (10^{-9})$~\cite{FCC:2018evy}.

\section{Projected sensitivity at the HL-LHC}\label{sec:proj}

\begin{figure}[!t]
    \centering
    \includegraphics[scale=0.4]{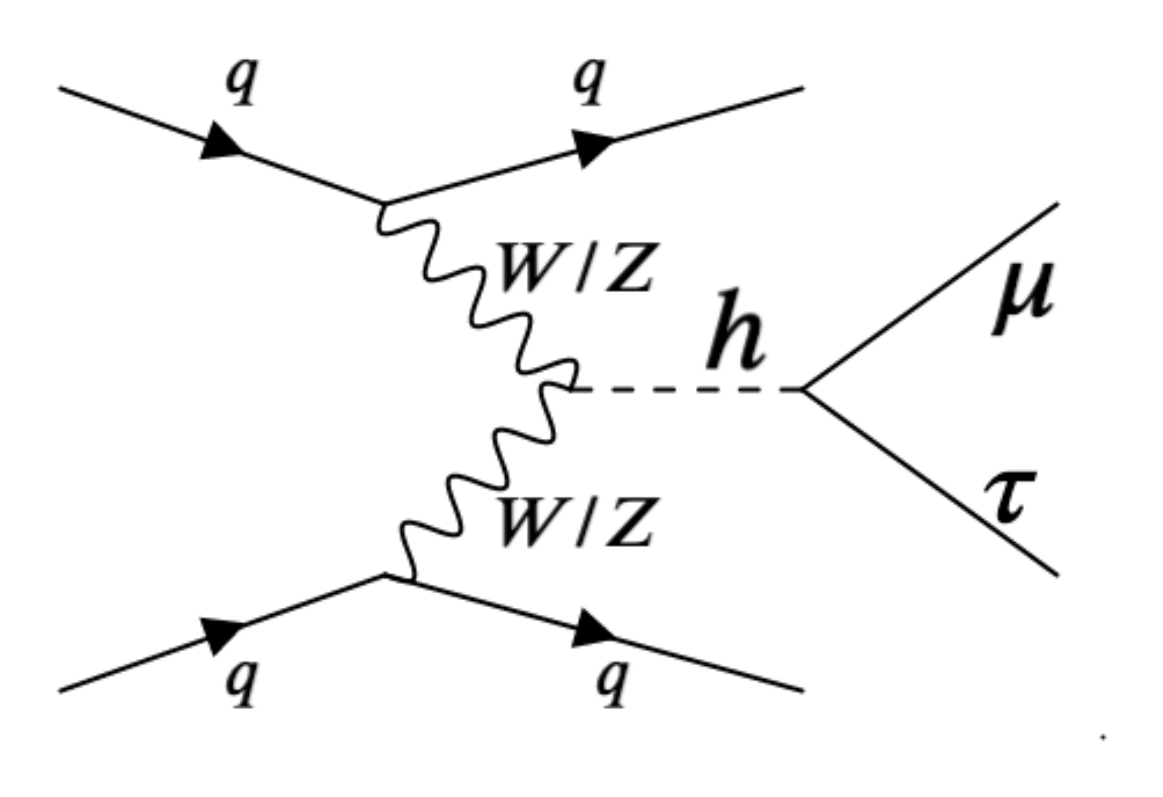}
    \caption{Leading order Feynman diagram for Higgs production in association with two jets in the vector boson fusion mode.}
    \label{fig:vbf_feynman}
\end{figure}

We study the projected sensitivity for the LFV couplings of the 125 GeV SM Higgs boson at the high luminosity LHC ($\sqrt{s}=14~\mathrm{TeV}$, $\mathrm{L}=3~\mathrm{ab^{-1}}$) through searches in VBF Higgs production channel 
\begin{equation}
    pp \to hjj \to (h \to \mu\tau)jj,
\end{equation}
where $\tau$ leptons can decay leptonically $\tau_{e} \to e + \nu_{e} + \nu_{\tau}$\footnote{Since we have a muon final state from $h\to \mu\tau$, we do not consider the tau decay into muon.} or hadronically $\tau_{h} \to \mathrm{hadrons}+\nu_{\tau}$. Representative Feynman diagram for the signal is shown in Fig.~\ref{fig:vbf_feynman}. 

Both leptonic and hadronic decay modes of the $\tau$ leptons are considered in the present analysis. In the $\mu\tau_{e}$ channel, we require exactly one isolated muon and one oppositely charged isolated electron in the final state, with $p_{T,\mu/e} > 15~\mathrm{GeV}$ and $|\eta| < 4.0$. Likewise in the $\mu\tau_{h}$ channel, we require exactly one isolated muon with the aforesaid trigger cuts and one oppositely charged $\tau$ tagged jet with $p_{T,\tau_{h}} > 25~\mathrm{GeV}$ and $|\eta| < 4.5$. Both channels are also required to have at least two light flavored jets ($j$) with $p_{T} > 30~\mathrm{GeV}$ and $|\eta| < 4.5$. The $\eta$ cuts are less stringent than the recent ATLAS~\cite{ATLAS:2019pmk} and CMS~\cite{CMS:2021rsq} studies in the same channel due to larger $\eta$ coverage at the HL-LHC~\cite{ZurbanoFernandez:2020cco}.

The important backgrounds are $Z+\mathrm{jets}$, $t\bar{t}$, multi-jet and $W$+jets with jets misidentified as leptons or $\tau$-tagged jets, and  $VV+\mathrm{jets}$, while sub-leading contributions can arise from single Higgs production in the VBF and gluon-gluon-fusion~(ggF) channel with $h \to \tau\tau$ and $h \to W^{+}W^{-}$. Furthermore, ggF mediated single Higgs production with LFV Higgs decay can also contribute to the VBF signal~\cite{DelDuca:2001eu,ATLAS:2019pmk,Barman:2020ulr}. We generate the signal and background events with $\texttt{MG5\_aMC@NLO}$~\cite{Alwall:2014hca, Artoisenet:2012st, Alwall:2007fs} at the leading order with $\sqrt{s}=14~$TeV. Signal and background events are simulated with generator-level cuts on the transverse momentum $p_{T}$ and pseudorapidity $\eta$ for the light flavored jets and leptons, $p_{T,j/\ell} > 10~$GeV and $|\eta_{j\ell}|<5.0$. A minimum threshold on the dijet invariant mass, $m_{jj} > 300~$GeV, is applied at the generator level for the background events. \texttt{Pythia8}~\cite{Sjostrand:2007gs} is used for parton showering and hadronization. The detector response is simulated using \texttt{Delphes-3.5.0}~\cite{deFavereau:2013fsa} with the default HL-LHC detector card~\cite{Cepeda:2650162, HLLHC_card}.

We closely follow the analysis strategy in a recent ATLAS study for the $\sqrt s=13$ TeV LHC~\cite{ATLAS:2019pmk}. In the $\mu\tau_{e}$ channel, the leading and subleading $p_{T}$ leptons, $\ell_{1}$ and $\ell_{2}$, respectively, are required to satisfy $p_{T,\ell_{1}} > 45~\mathrm{GeV}$ and $p_{T,\ell_{2}} > 15~\mathrm{GeV}$. The asymmetric $p_{T}$ cuts are used to suppress the $h \to \tau^{+}\tau^{-}$ background. We also veto events containing a third isolated electron or muon to reduce the diboson background. Similarly, in the $\mu\tau_{h}$ channel,  we require $p_{T,\mu} > 30~\mathrm{GeV}$ and $p_{T,\tau_{h}} > 45~\mathrm{GeV}$. We also require the sum of cosine of azimuthal angle differences among the $\{\mu,\slashed{E}_{T}\}$ and $\{\tau_{h},\slashed{E}_{T}\}$ pairs, $\sum_{i=\mu,\tau_{h}} \cos\Delta \Phi(i,\slashed{E}_{T})$, to be greater than $> -0.35$ in order to reduce the $W+\mathrm{jets}$ background. Furthermore, an upper limit is imposed on the pseudorapidity difference between $\mu$ and $\tau_{h}$, $|\Delta \eta(\mu,\tau_{h})| < 2.0$ to suppress the contributions arising from multi-jet backgrounds~\cite{ATLAS:2019pmk}. To reduce the massive $t\bar{t}$ background, events containing any $b$ tagged jet with $p_{T} > 25~\mathrm{GeV}$ and $|\eta| < 4.0$ are vetoed in both channels.

\begin{figure}[t!]
    \centering
    \includegraphics[scale=0.35]{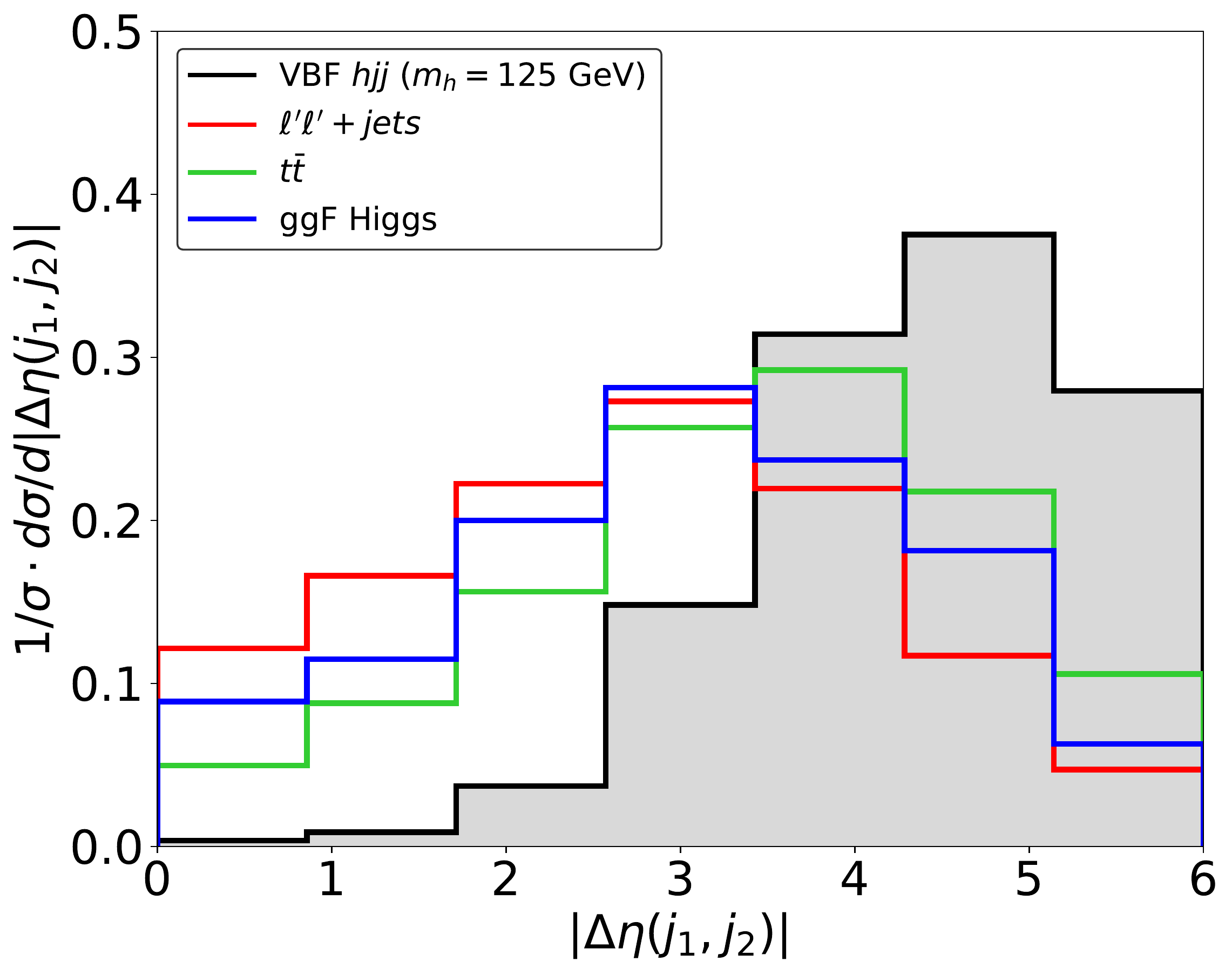}
    \caption{Distributions for the difference of pseudorapidities for the  VBF tagged jets $|\Delta \eta(j_1,j_2)|$ at the detector level for the $\ell^{\prime}\ell^{\prime}+jets$~(red) and $t\bar{t}$~(green) backgrounds, $gg \to h_{125} $~(blue) and VBF~(black) signal events at $m_{h} = 125~$GeV. The events satisfy the basic selection cuts for the $\mu\tau_{e}$ channel in Table~\ref{tab:selection_cuts}, $p_{T,j_1}>40~$GeV, $p_{T,j_2}>30~$GeV and $m_{j_1j_2}>350~$GeV. We assume $\sqrt{s}=14~$TeV LHC with $\mathcal{L}=3~\rm{ab^{-1}}$.}
    \label{fig:detajj_sm125}
\end{figure}

The next objective is to identify and reconstruct the VBF topology. The leading and subleading $p_{T}$ light jets $j_{1}$ and $j_{2}$, respectively, are tagged as VBF jets provided they satisfy $p_{T,j_{1}~(j_2)} > 40~\mathrm{GeV}$~(30~GeV). The VBF jets feature a large invariant mass $m_{j_1 j_2}$. Correspondingly, we impose a minimum threshold of $m_{j_1j_2} > 350~\mathrm{GeV}$. Furthermore, the VBF jets are mostly produced back to back in the forward regions of the detector, thus having $\eta_{j_1}\cdot \eta_{j_2} < 0$, and are characterized by a large pseudorapidity difference $|\Delta \eta(j_1,j_2)|$. For illustrative purposes, we present the distributions for $|\Delta \eta(j_1,j_2)|$ in Fig.~\ref{fig:detajj_sm125}. Since the VBF jets are mostly populated at large $|\Delta \eta(j_1,j_2)|$ regions, we require the events to satisfy $|\Delta \eta(j_1,j_2)| > 2$ at the event selection stage and also use $|\Delta \eta(j_1,j_2)|$ as a training observable in the multivariate analysis. The VBF topology also features reduced jet activity in the central region. Therefore, events with additional light jets~($j_{3}$) in the central region are vetoed by imposing $|\eta_{j_{3}} - (\eta_{j_1} + \eta_{j_2})/2| > 1.0$.  The event selection cuts for the $\mu\tau_{e}$ and $\mu\tau_{h}$ channels are summarized in Table~\ref{tab:selection_cuts}.   

\begin{table}[!t]
    \centering\scalebox{0.9}{
    \begin{tabular}{|c|c|c|} \hline
        & $\mu\tau_{e}$ & $\mu\tau_{h}$ \\ \hline \hline
        & $n_{\mu} = 1$, $n_{e} = 1$ & $n_{\mu} = 1$, $n_{\tau_{h}} = 1$  \\ 
        & ($e^{+}\mu^{-}$/$e^{-}\mu^{+}$)  & ($\tau_{h}^{+}\mu^{-}$/$\tau_{h}^{-}\mu^{+}$) \\
        Basic & $p_{T,\ell_{1}} > 45~\mathrm{GeV}$ & $p_{T,\tau_{h}} > 45~\mathrm{GeV}$ \\
        Selection & $p_{T,\ell_{2}} > 15~\mathrm{GeV}$ & $p_{T,\mu} > 30~\mathrm{GeV}$ \\ 
        &  & $\sum\limits_{i=\mu,\tau_{h}} \cos\Delta \Phi(i,\slashed{E}_{T}) > -0.35$ \\ 
        &  & $|\Delta \eta(\mu,\tau_{h})| < 2.0$ \\ \cline{2-3}
        & \multicolumn{2}{c|}{$b$ jet veto if $p_{T} > 25~\mathrm{GeV}$ and $|\eta| < 4.0$}\\ \hline
        & \multicolumn{2}{c|}{$p_{T,j_{1}} > 40~\mathrm{GeV}$, $p_{T,j_{2}} > 30~\mathrm{GeV}$,} \\
        VBF & \multicolumn{2}{c|}{$|\Delta \eta(j_1,j_2)| > 2$, $\eta_{j_1}\cdot \eta_{j_2} < 0$, } \\
        topology & \multicolumn{2}{c|}{$m_{j_1j_2} > 350~\mathrm{GeV}$} \\
        & \multicolumn{2}{c|}{$|\eta_{j_{3}} - (\eta_{j_1} + \eta_{j_2})/2| > 1.0$} \\ \hline
    \end{tabular}}
    \caption{Summary of event selection cuts for the $\mu\tau_{e}$ and $\mu\tau_{h}$ channels.}
    \label{tab:selection_cuts}
\end{table}

\begin{figure*}[!htb]
    \centering
    \includegraphics[width=0.495\textwidth]{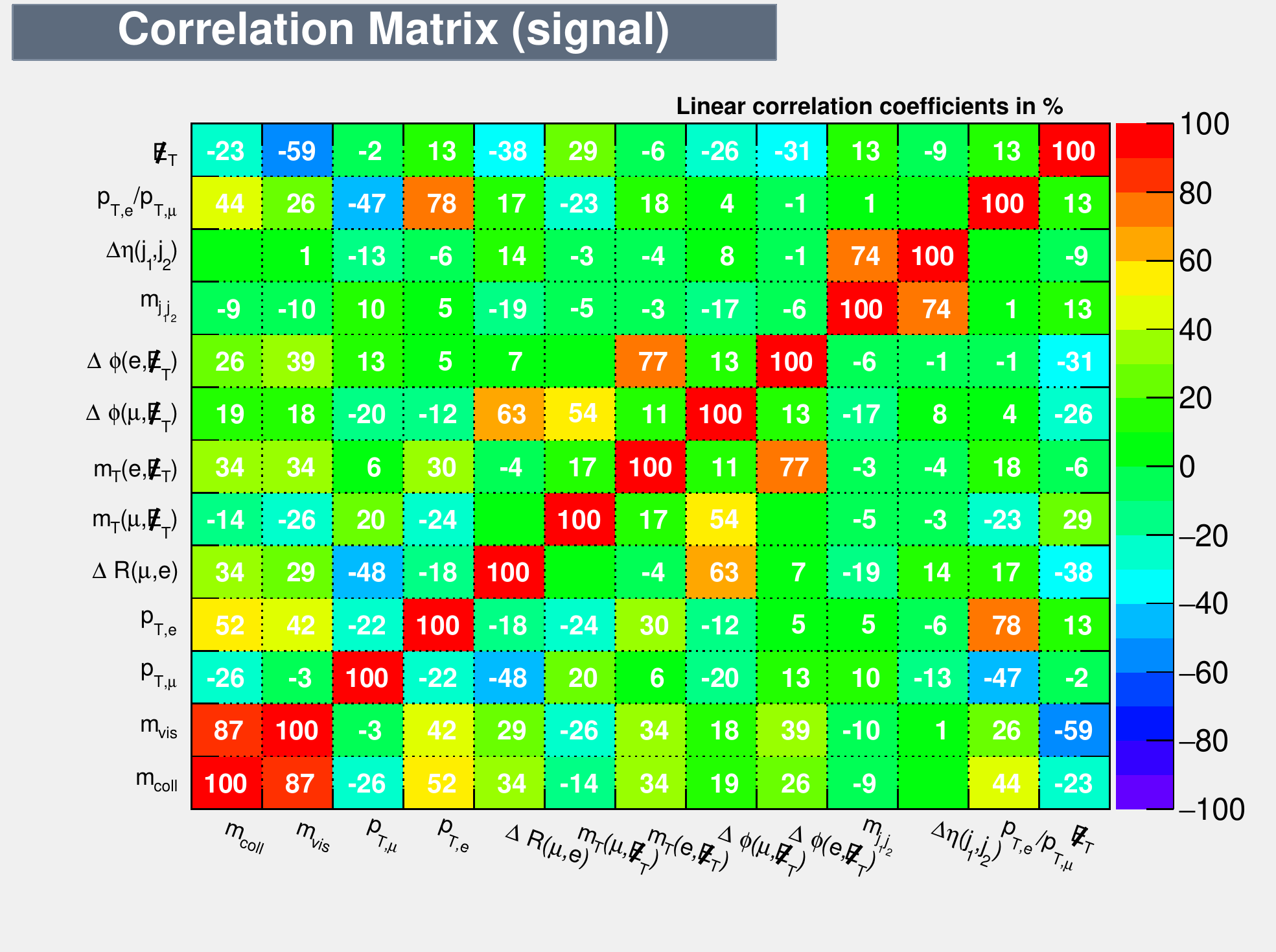}
    \includegraphics[width=0.495\textwidth]{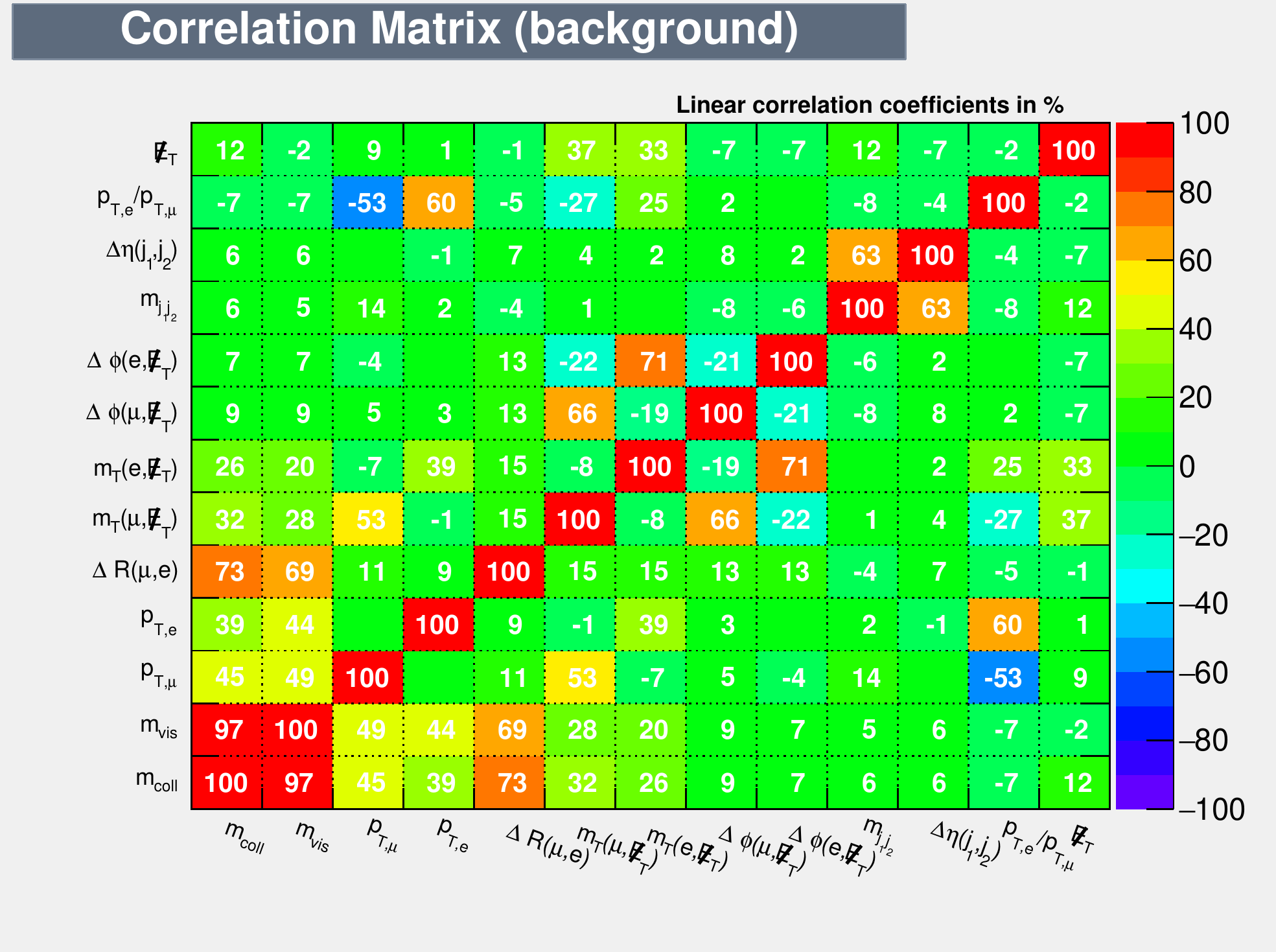}
    \caption{Linear correlation coefficient matrix for input variables to the BDT for signal and background events in the $\mu\tau_{e}$ channel.}
    \label{fig:corr_mue}
\end{figure*}

We perform a multivariate analysis using the Boosted Decision Tree (BDT) algorithm to discriminate the VBF signal from SM backgrounds. The kinematic observables used to perform the multivariate analysis are shown in Table~\ref{tab:training_obs}. 
\begin{table}[!t]
    \centering
    \begin{tabular}{|c|c|} \hline
        $\mu\tau_{e}$ & $\mu\tau_{h}$ \\ \hline
        $p_{T,e}, m_{T}(e,\slashed{E}_{T})$ & $p_{T,\tau_{h}}, m_{T}(\tau_{h},\slashed{E}_{T}),$ \\
        $\Delta R(\mu,e), \Delta \phi(e,\slashed{E}_{T})$ & $\Delta R(\mu,\tau_{h}), \Delta \phi(\tau_{h},\slashed{E}_{T})$\\
        $p_{T,e}/p_{T,\mu}, \eta_{e}, \phi_{e}$ & $ \eta_{\tau_{h}}, \phi_{\tau_{h}}, \sum\limits_{i=\mu,\tau_{h}} \cos\Delta \Phi(i,\slashed{E}_{T}),$ \\ \hline 
        \multicolumn{2}{|c|}{$p_{T,\mu}, m_{T}(\mu,\slashed{E}_{T}), \Delta \phi(\mu,\slashed{E}_{T}), \eta_{\mu}, \phi_{\mu}, \phi_{\slashed{E}_{T}},$} \\
        \multicolumn{2}{|c|}{$m_{\rm coll}, m_{\rm vis}, m_{j_1j_2}, \Delta\eta(j_1,j_2), \slashed{E}_{T} $} \\ \hline
    \end{tabular}
    \caption{Input observables used in the BDT analysis for the $\mu\tau_{e}$ and $\mu\tau_{h}$ channels.}
    \label{tab:training_obs}
\end{table}

Here, $m_{T}(\alpha,\slashed{E}_{T})~(\alpha=\mu,e)$ is the transverse mass for the lepton $\alpha$ and $\slashed{E}_{T}$ pair, defined as $m_{T}(\alpha,\slashed{E}_{T}) = \sqrt{2p_{T,\alpha}\slashed{E}_{T}(1-\cos\Delta\phi(\alpha,\slashed{E}_{T}))}$, $\Delta \phi(\beta,\slashed{E}_{T})$~($\beta = \alpha, \tau_{h}$) is the difference in azimuthal angles for object $\beta$ and $\slashed{E}_{T}$, and $m_{\mathrm{vis}}$ is the visible invariant mass for the $h \to \mu(\tau \to e \nu_{\tau}\nu_{e})$ system. The full reconstruction of the Higgs boson is challenging since the $\tau$ decay is associated with missing energy. Several techniques have been developed to reconstruct the resonant invariant mass in such cases (cf.~Refs.~\cite{Ellis:1987xu,Elagin:2010aw,Hagiwara:2016zqz} and references therein). In this analysis, we adopt the collinear mass approximation technique which is rooted on two important assumptions that the visible and invisible decay products of $\tau$ are collinear, and the only source of missing energy in the system is the neutrinos from $\tau$ decay. Following Ref.~\cite{Elagin:2010aw}, the collinear mass is computed as $m_{\mathrm{coll}} = m_{\mathrm{vis}}/\sqrt{x_{\tau_{e}}}$, where $x_{\tau_{e}} = p_{T,\tau_{e}}/(p_{T,\tau_{e}} + \slashed{E}_{T})$. The rest of the observables in Table~\ref{tab:training_obs} have their usual meaning. 

\begin{figure*}[!htb]
    \centering
    \includegraphics[width=0.495\textwidth]{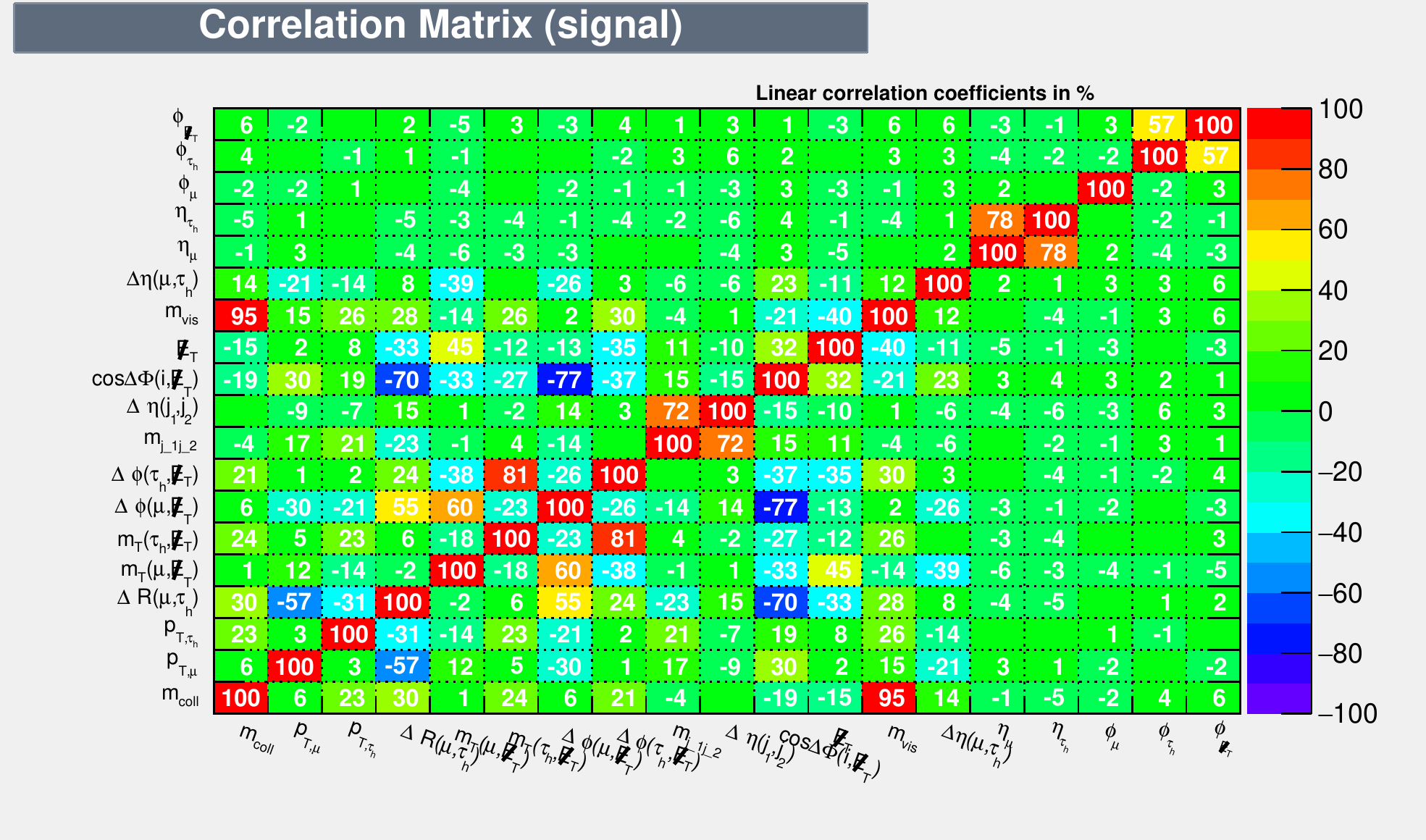}
    \includegraphics[width=0.495\textwidth]{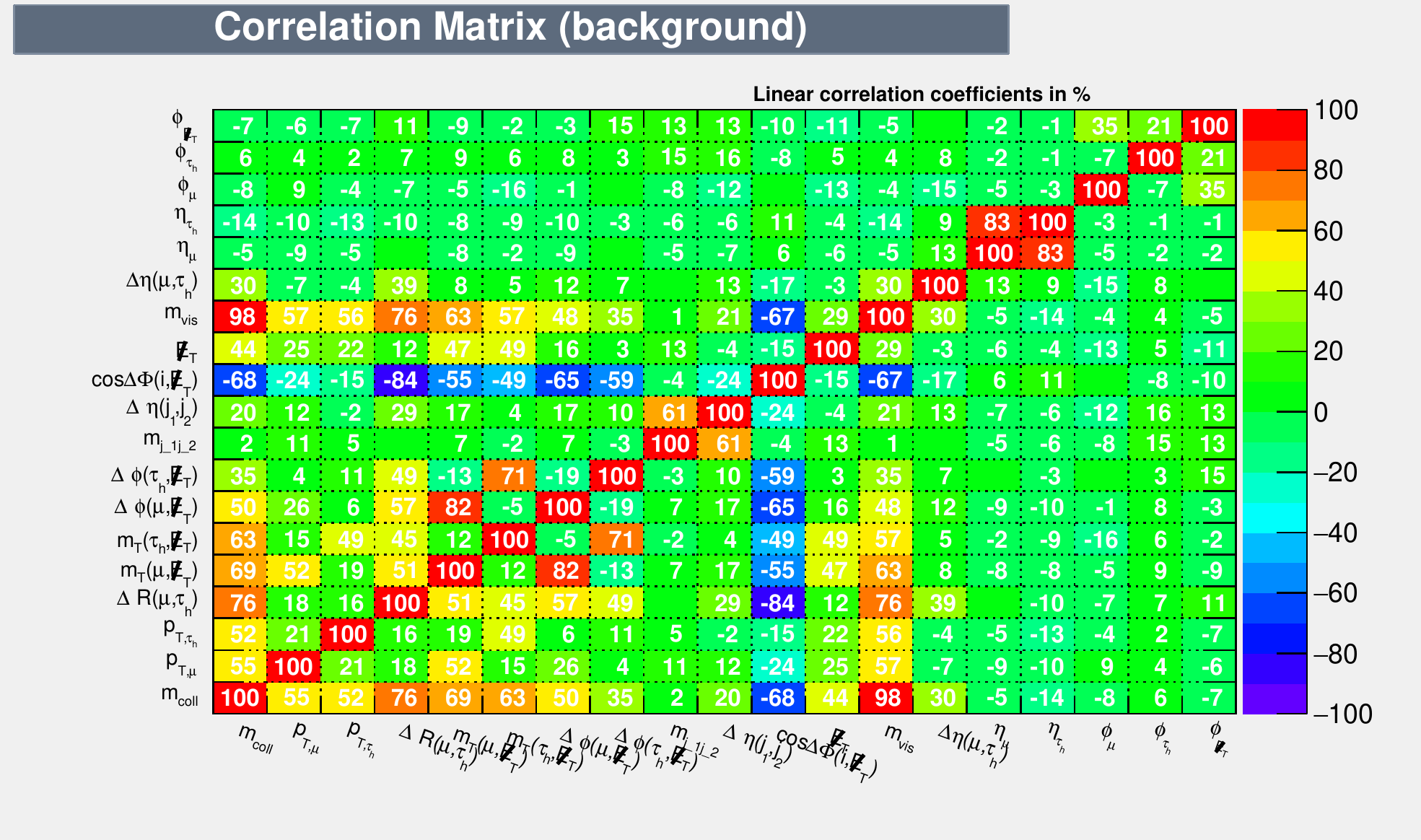}
    \caption{Linear correlation coefficient matrix for input variables to the BDT for signal and background events in the $\mu\tau_{h}$ channel.}
    \label{fig:corr_muj}
\end{figure*}

In Figs.~\ref{fig:corr_mue} and \ref{fig:corr_muj}, we show the correlation among the input variables for signal and background events in the $\mu\tau_{e}$ and $\mu\tau_{h}$ channels, respectively. Due to high correlation ($\gtrsim 80\%$) between $m_{\rm coll}$ and $m_{\rm vis}$, the latter is not considered in the training. Similarly, in the $\mu\tau_{h}$ channel, the highly correlated observable pairs are $\{m_{\rm vis},m_{\rm coll}\}$ and $\{\Delta \phi(\tau_{h},\slashed{E}_{T}),m_{T}(\tau_{h},\slashed{E}_{T})\}$, among which, we include only the latter two observables in the training. The kinematic observables from Tab.~\ref{tab:training_obs} with the highest ranks in the BDT analysis are,
\begin{equation}
\begin{aligned}
     \mu\tau_{e} &: m_{\rm coll}, m_{T}(\mu/e,\slashed{E}_{T}), \Delta R(\mu,e), \slashed{E}_{T}, \Delta \Phi(\mu,\slashed{E}_{T}) \\  \mu\tau_{h} &: m_{\rm coll}, m_{T}(\tau_{h},\slashed{E}_{T}), \Delta \Phi(\mu,\slashed{E}_{T}), \Delta \Phi(\mu,\slashed{E}_{T}), \slashed{E}_{T}, \phi_{\mu}. \nonumber 
\end{aligned}
\end{equation}

\begin{figure*}[!t]
    \centering
    \includegraphics[width=0.8\textwidth]{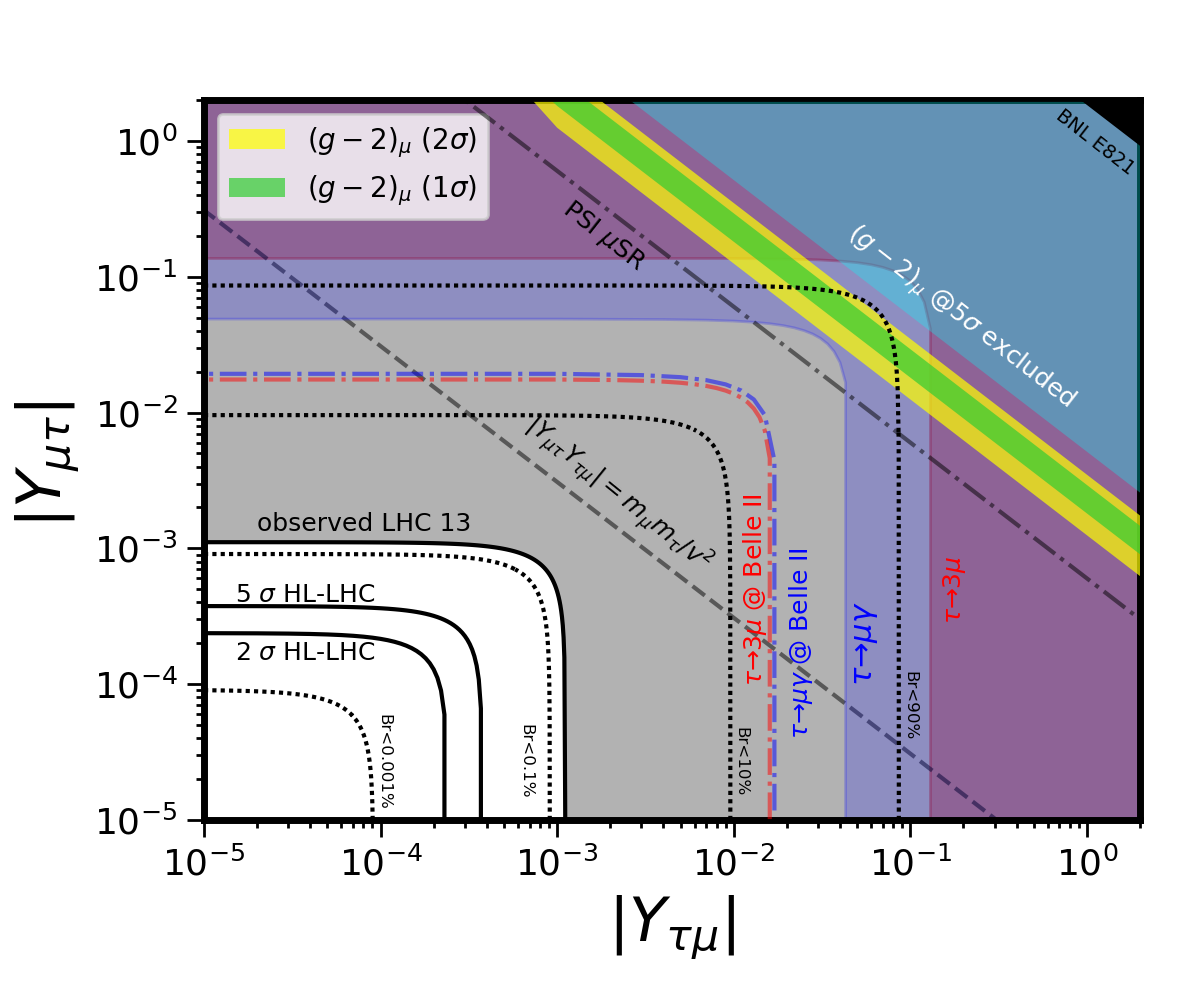}
    \caption{Projected upper limits on the off-diagonal Yukawa couplings $|Y_{\mu\tau}|$ and $|Y_{\tau\mu}|$ from direct searches in the combined $\mu\tau_e+\mu\tau_{h}$ channel at the HL-LHC are represented as solid black lines (see also Table~\ref{tab:sm_cut_flow}). The current limit from 13 TeV LHC data~\cite{CMS:2021rsq} is shown by the black-shaded region. The red (blue) shaded region represents the exclusion from LFV decays $\tau\to 3 \mu$~\cite{Hayasaka:2010np}  ($\tau \to \mu \gamma$~\cite{BaBar:2009hkt}), whereas the red and blue dash-dotted lines are the corresponding future projected sensitivities at Belle II~\cite{Belle-II:2022cgf}. The green (yellow) band represents the preferred parameter space to accommodate the $(g-2)_\mu$ anomaly~\cite{Muong-2:2021ojo} at $1\sigma$ ($2\sigma$). The black and cyan shaded regions at the top right corner respectively represent the exclusions from $\mu$EDM~\cite{Muong-2:2008ebm} (only if $\Im(Y_{\mu\tau}Y_{\tau\mu})\neq 0$) and $(g-2)_\mu$~\cite{Muong-2:2021ojo} at $5\sigma$, whereas the black dash-dotted diagonal line corresponds to the future sensitivity reach of $\mu$EDM~\cite{Adelmann:2021udj} for $\Im(Y_{\mu\tau}Y_{\tau\mu})$. The black dashed diagonal line in the middle is the theoretical naturalness limit $Y_{\tau\mu}Y_{\mu\tau}=m_\mu m_\tau/v^2$~\cite{Cheng:1987rs}. The dotted contours correspond to different values of the $h\to \mu\tau$ branching ratios (0.001\%, 0.1\%, 10\% and 90\%).   }
    \label{fig:LFV_Sm_Higgs}
\end{figure*}

In the $\mu\tau_{e}$ channel, the contribution from multi-jet events to the total background is expected to be suppressed since the probability for two jets being misidentified as isolated leptons with different flavors is relatively small. Previous searches performed in the aforesaid channel using the LHC Run-II data collected at $\mathcal{L}=36.1~\rm{fb}^{-1}$~\cite{ATLAS:2019pmk} indicate that the relative contribution from misidentified backgrounds, which include $W$+jets and multi-jet processes, is around $5\%$ of the total background. Therefore, we ignore the misidentified backgrounds in our analysis for the $\mu\tau_{e}$ channel. However, in the $\mu\tau_{h}$ channel, misidentified events have a non-trivial contribution. The analysis for the VBF $\mu\tau_{h}$ channel in Ref.~\cite{ATLAS:2019pmk} indicates the contribution from misidentified backgrounds to be roughly $25\%$, and hence, cannot be neglected. However, simulating the QCD multi-jet and $W$+jets background relevant for the $\mu\tau_{h}$ channel is very challenging, and is mostly estimated through data-driven techniques which is beyond the scope of the present work. The simulated background used in our BDT analysis constitutes roughly $75\%$ of the full background. Accordingly, the ratio of the left-out misidentified background and the simulated background contribution to the total background is around $1/3$. In order to account for the underestimated background rate, we adopt a simplistic approach of compounding the total yield from other simulated background processes after the optimized BDT analysis by a factor of $(1+1/3) = 4/3$. It must be noted that the inclusion of misidentified events in the BDT optimization itself might lead to marginally different sensitivity. As such, our results for the $\mu\tau_{h}$ channel are only conservative projections, and must be viewed as such.  

\begin{table}[!t]
    \centering
    \begin{tabular}{|c|c|c|c|c|c|}\hline
    \multirow{2}{*}{Channel} & \multicolumn{2}{|c|}{Signal eff.} & \multirow{2}{*}{Bkg.} & \multicolumn{2}{c|}{${\rm Br}(h\to\mu\tau)~(2\sigma)$ U.L.} \\ \cline{2-3}\cline{5-6}
     &  VBF & ggF &  & $\kappa = 0$ & $\kappa = 5\%$ \\ \hline 
    $\mu\tau_{e}$ & $1.13\times 10^{-2}$  & $4.5\times 10^{-4}$ & 3192 & $5.4\times10^{-4}$ & $1.6\times 10^{-3}$\\ \hline
    $\mu\tau_{h}$ & $9.2\times 10^{-3}$ & $3.0\times 10^{-4}$ & 403 & $2.5\times 10^{-4}$ & $3.5\times 10^{-4}$ \\ \hline 
    \multicolumn{4}{|c|}{$\mu\tau_{e}$ + $\mu\tau_{h}$ combined} & $3.2\times 10^{-4}$ & $1.0\times 10^{-3}$ \\ \hline 
    \end{tabular}
    \caption{Signal efficiency and background yields from searches for LFV decays of Higgs boson $pp \to (h\to\mu\tau)jj$ in the $\mu\tau_{e}$ and $\mu\tau_{h}$ channels at the $\sqrt{s}=14~$TeV LHC with $\mathcal{L}=3~\rm{ab}^{-1}$. Projected upper limits (U.L.) on the LFV branching ratios are also shown for null and $5\%$ systematic uncertainty.}
    \label{tab:sm_cut_flow}
\end{table}

The signal efficiency~($\epsilon_{S}$) and background yields~($B$) at the HL-LHC from the BDT analysis are shown in Table~\ref{tab:sm_cut_flow}. Assuming SM rates for ggF and VBF Higgs production at $\sqrt{s}=14~$TeV, $\sigma_{h_{\rm SM}}^{\rm ggF} = 49.5~$pb~[computed at next-to-next-to-leading-order~(NNLO) QCD $+$ next-to-next-to-leading-logarithm~(NNLL) QCD $+$ next-to-leading-order~(NLO) electroweak~(EW)] and $\sigma_{h_{\rm SM}}^{\rm VBF}= 4.2~$pb~(computed at NNLO QCD $+$ NLO EW)~\cite{LHCHiggsCrossSectionWorkingGroup:2013rie},  the relative contribution from the ggF signal to the total signal rate is roughly $30\%$ and $27\%$ in the $\mu\tau_{e}$ and $\mu\tau_{h}$ channels, respectively. We translate the results into projected upper limits on the branching ratio for LFV Higgs decays in Table~\ref{tab:sm_cut_flow} using,
\begin{equation}
    {\rm Br}(h \to \mu\tau) = \frac{n_{S} \cdot  \sqrt{B + (\kappa\cdot B)^{2}}}{\mathcal{L}\times (\sigma_{h_{\rm SM}}^{\rm ggF}\cdot \epsilon_{S}^{\rm ggF} + \sigma_{h_{\rm SM}}^{\rm VBF}\cdot \epsilon_{S}^{\rm VBF})},
\label{eq:ul_func}
\end{equation}
where $n_{S}$ is the standard deviation from background, and $\kappa$ is the systematic uncertainty. We observe that the HL-LHC would be able to probe Higgs LFV decays in the VBF Higgs production channel up to ${\rm Br}(h \to \mu\tau) \sim 0.032\%$ at $2\sigma$ in a quasi-ideal scenario with no systematic uncertainties. The limit weakens to ${\rm Br}(h \to \mu\tau) \sim 0.1\%$ upon considering $5\%$ systematic uncertainty. The current CMS limit on ${\rm Br}(h \to \mu\tau)$ from combined searches in $\mu\tau_{e}$, $\mu\tau_{h}$, $e\tau_{\mu}$ and $e\tau_{h}$ channels with ggF and VBF Higgs production, using the LHC Run-II data collected at $\mathcal{L}=137~\rm{fb}^{-1}$, is ${\rm Br}(h \to \mu\tau) < 0.15\%$ at $95\%$ confidence level (CL)~\cite{CMS:2021rsq}. The corresponding ATLAS limit is slightly weaker at ${\rm Br}(h \to \mu\tau) < 0.18\%$~\cite{ATLAS:2022vjy}. The projected reach at the HL-LHC from combined searches in the aforesaid channels with both ggF and VBF Higgs production has been estimated at ${\rm Br}(h \to \mu\tau) \lesssim 0.05\%$ in Ref.~\cite{Davidek:2020gbw} through a luminosity scaling of the current limits, and is roughly $2$ times stronger than the projected sensitivity derived in the present analysis for $\kappa=5\%$. Note that the search potential for the VBF channel alone is complementary to the combined projected reach.

The LFV Higgs branching fraction is related to the off-diagonal Yukawa couplings,
\begin{eqnarray}
     |Y_{\mu\tau}|^{2}+|Y_{\tau\mu}|^{2} = \frac{8\pi}{m_{h}}\frac{{\rm Br}(h \to \mu\tau)}{1-{\rm Br}(h \to \mu\tau)}\Gamma_{h},
\label{eq:sm_br_ymutau}
\end{eqnarray}
where, $\Gamma_{h} = 4.07~$MeV~\cite{LHCHiggsCrossSectionWorkingGroup:2013rie} is the decay width for the SM Higgs boson. Using Eq.~\eqref{eq:sm_br_ymutau}, upper limits on the LFV Higgs branching ratio can be translated into limits for $Y_{\mu\tau}$ and $Y_{\tau\mu}$, $\sqrt{|Y_{\mu\tau}|^{2} + |Y_{\tau\mu}|^{2}} < 5.1\times 10^{-4}$. We present these projected upper limits in the plane of $|Y_{\mu\tau}|$ and $|Y_{\tau\mu}|$ in Fig.~\ref{fig:LFV_Sm_Higgs} as solid-black lines.  The current LHC exclusion region at $95\%$ CL, $\sqrt{Y_{\mu\tau}^{2} + Y_{\tau\mu}^{2}} < 1.1\times10^{-3}$~\cite{CMS:2021rsq,ATLAS-CONF-2022-060}, is displayed in shaded black. Other low-energy constraints and future projections discussed in Sec.~\ref{sec:lowenergy} are also shown for comparison. The black-dashed line is the naturalness bound given in Eq.~\eqref{eq:natural}. It is clear that the HL-LHC projected sensitivities surpass the future low-energy constraints from $\tau\to \mu\gamma$ and $\tau\to 3\mu$ expected at Belle II~\cite{Belle-II:2018jsg}. 

We also find from Fig.~\ref{fig:LFV_Sm_Higgs} that the current $(g-2)_\mu$ anomaly cannot be accommodated by LFV couplings of the SM Higgs boson, as the preferred region (green/yellow-shaded for $1\sigma/2\sigma$) is already excluded by both $\tau\to \mu\gamma$ constraint, as well as by the 13 TeV LHC upper limit on ${\rm Br}(h\to \mu\tau)$. It should however be noted that the significance of this anomaly has recently diminished, because of recent lattice developments~\cite{Borsanyi:2020mff, Ce:2022kxy,Alexandrou:2022amy}. So it remains to be seen whether we need any BSM physics in the muon $g-2$ sector.

\section{BSM Higgs LFV decay at the HL-LHC}
\label{sec:BSM}

In this section, we further extend the analysis strategy discussed in Sec.~\ref{sec:proj} to study the projected sensitivity for LFV decays of a generic BSM leptophilic Higgs bosons $H$ produced in the VBF mode, $ pp \to Hjj \to (H \to \mu\tau)jj$, at the HL-LHC. Because of its leptophilic nature, the contributions from ggF Higgs production to the signal are not taken into account. We perform the BDT analysis for five signal benchmarks with $m_{H} < m_{h_{\rm SM}}$, $m_{H} = 20, 40, 60, 80$ and 100~GeV, and five benchmarks with $m_{H} > m_{h_{\rm SM}}$, $m_{H} = 150, 200, 400, 600$ and 800~GeV. Analogous to the earlier analysis, the search is performed in both $\mu\tau_{h}$ and $\mu\tau_{e}$ channels using the kinematic observables iterated in Table~\ref{tab:training_obs}. The observables found useful for our BDT analysis are identified in Tables~\ref{tab:obs_rank_mutaue} and \ref{tab:obs_rank_mutauj} for the $\mu \tau_e$ and $\mu \tau_h$ channels, respectively.  

\begin{table}[!t]
    \centering
    \begin{tabular}{|c|c|c|c|c|c|c|c|c|c|c|c|} \hline 
     \multirow{2}{*}{$\mu\tau_{e}$} & \multicolumn{11}{c|}{$m_{H}$ [GeV]}\\ \cline{2-12}
       & 20 & 40 & 60 & 80 & 100 & 125 & 150 & 200 & 400 & 600 & 800 \\ \hline 
      $p_{T,e}$ & \ding{52} & $\bullet$ & $\bullet$ & $\bullet$ & $\bullet$ & $\bullet$ & $\bullet$ & $\bullet$ & $\bullet$ & $\bullet$ & \\
      $m_{T}(e,\slashed{E}_{T})$ & \ding{52} & \ding{52} & $\bullet$ & \ding{52} & $\bullet$ & \ding{52} & \ding{52} & \ding{52} & \ding{52} & \ding{52} & \ding{52} \\
      $\Delta R(\mu,e)$ & \ding{52} & \ding{52} & \ding{52} & \ding{52} & \ding{52} & \ding{52} & \ding{52} & \ding{52} & $\bullet$ & $\bullet$ & $\bullet$ \\
      $\Delta \phi(e,\slashed{E}_{T})$ &  & \ding{52}  & $\bullet$ & $\bullet$ &  \ding{52} & $\bullet$ & \ding{52} & $\bullet$ & $\bullet$ & \ding{52} & \ding{52} \\
      $p_{T,\mu}$ & $\bullet$ & $\bullet$ & \ding{52} & $\bullet$ & $\bullet$ & $\bullet$ & $\bullet$ & \ding{52} & \ding{52} & \ding{52} & \ding{52} \\
      $m_{T}(\mu,\slashed{E}_T)$ &  & $\bullet$ & $\bullet$ & $\bullet$ & $\bullet$ & \ding{52} & $\bullet$ & $\bullet$ & \ding{52} & \ding{52} & \ding{52} \\
      $\Delta \phi(\mu,\slashed{E}_{T})$ &  &  & \ding{52} & \ding{52} & \ding{52} & \ding{52} & \ding{52} & \ding{52} & \ding{52} & \ding{52} &  $\bullet$ \\
      $m_{\rm coll}$ &  & \ding{52} & \ding{52} & \ding{52} & \ding{52} & \ding{52} & \ding{52} & \ding{52} & \ding{52} & \ding{52} & \ding{52} \\
      $m_{j_2 j_2}$ & $\bullet$ & $\bullet$ & $\bullet$ & $\bullet$ & $\bullet$ & $\bullet$ & $\bullet$ & $\bullet$ & $\bullet$ & $\bullet$ & $\bullet$ \\
      $\Delta \eta(j_1,j_2)$ & \ding{52} & \ding{52} & \ding{52} & \ding{52} & \ding{52} & $\bullet$ & $\bullet$ & \ding{52} & \ding{52} & $\bullet$ & $\bullet$ \\
      $\slashed{E}_T$ & \ding{52} & \ding{52} & \ding{52} & \ding{52} & \ding{52} & \ding{52} & \ding{52}  & $\bullet$ & $\bullet$ & $\bullet$ &  \ding{52} \\ 
      $p_{T,e}/p_{T,\mu}$ & \ding{52} & $\bullet$ & $\bullet$ & $\bullet$ & $\bullet$ &  &  &  &  &  & \\\hline
    \end{tabular}
    \caption{Kinematic observables used in the BDT analysis for the $\mu\tau_{e}$ channel with different BSM Higgs boson masses. The used observables are marked with a  \ding{52}~(shows the six observables with the highest rank) or a $\bullet$.}
    \label{tab:obs_rank_mutaue}
\end{table}

\begin{table}[!t]
    \centering
    \begin{tabular}{|c|c|c|c|c|c|c|c|c|c|c|c|} \hline 
     \multirow{2}{*}{$\mu\tau_{h}$} & \multicolumn{11}{c|}{$m_{H}$ [GeV]}\\ \cline{2-12}
       & 20 & 40 & 60 & 80 & 100 & 125 & 150 & 200 & 400 & 600 & 800 \\ \hline
      $p_{T,\tau_{h}}$ & $\bullet$ & $\bullet$ & $\bullet$ & $\bullet$ & $\bullet$ & $\bullet$ & $\bullet$ & $\bullet$ & $\bullet$ & \ding{52} & \ding{52} \\
      $m_{T}(\tau_{h},\slashed{E}_{T})$ & $\bullet$ & $\bullet$ & $\bullet$ & $\bullet$ & $\bullet$ & \ding{52} & \ding{52} & \ding{52} & \ding{52} & \ding{52} & \ding{52} \\
      $\Delta R(\mu,\tau_{h})$ &  & \ding{52} & $\bullet$ & $\bullet$ & $\bullet$ & $\bullet$ & $\bullet$ & $\bullet$ & $\bullet$ & $\bullet$ & $\bullet$ \\
      $\eta_{\tau_{h}}$ &  & $\bullet$ & $\bullet$ & $\bullet$ & $\bullet$ & $\bullet$ & $\bullet$ & $\bullet$ & $\bullet$ & $\bullet$ & $\bullet$ \\
      $\phi_{\tau_{h}}$ & \ding{52} & $\bullet$ & $\bullet$ & \ding{52} & \ding{52} & $\bullet$ & $\bullet$ & $\bullet$ & $\bullet$ & $\bullet$ & $\bullet$ \\
      $p_{T,\mu}$ & $\bullet$ & $\bullet$ & $\bullet$ & $\bullet$ & $\bullet$ & $\bullet$ & $\bullet$ & $\bullet$ & \ding{52} & \ding{52} & \ding{52}\\
      $m_{T}(\mu,\slashed{E}_{T})$ & $\bullet$ & $\bullet$ & $\bullet$ & $\bullet$ & $\bullet$ & $\bullet$ & $\bullet$ & \ding{52} & \ding{52} & \ding{52} & \ding{52} \\
      $\Delta \phi(\mu,\slashed{E}_{T})$ & $\bullet$ & $\bullet$ & $\bullet$ & $\bullet$ & $\bullet$ & \ding{52} & \ding{52} & \ding{52} & \ding{52} & \ding{52} & \ding{52} \\
      $\eta_{\mu}$ & \ding{52} & \ding{52} & \ding{52} & $\bullet$ & $\bullet$ & $\bullet$ & $\bullet$ & $\bullet$ & $\bullet$ & $\bullet$ & $\bullet$ \\
      $\phi_{\mu}$ & $\bullet$ & $\bullet$ & $\bullet$ & \ding{52} & \ding{52} & \ding{52} & \ding{52}  &  $\bullet$ & $\bullet$ & $\bullet$ & $\bullet$ \\
      $\phi_{\slashed{E}_{T}}$ & \ding{52} & $\bullet$ & \ding{52} & $\bullet$ & $\bullet$ & $\bullet$ & $\bullet$ & $\bullet$ & $\bullet$ & $\bullet$ & $\bullet$ \\
      $m_{\rm coll}$  & \ding{52} & \ding{52} & \ding{52} & \ding{52} & \ding{52} &  \ding{52} & \ding{52} & \ding{52} & \ding{52} & \ding{52} & \ding{52} \\
      $m_{j_1 j_2}$ & $\bullet$ & $\bullet$ & $\bullet$ & $\bullet$ & $\bullet$ & $\bullet$ & $\bullet$ & $\bullet$ & $\bullet$ & $\bullet$ & $\bullet$ \\
      $\Delta \eta(j_1,j_2)$ & $\bullet$ & \ding{52} & \ding{52} & \ding{52} & \ding{52} & $\bullet$ & $\bullet$ & $\bullet$ & $\bullet$ & $\bullet$ & $\bullet$ \\
      $\slashed{E}_{T}$ & \ding{52} & \ding{52} & \ding{52} & \ding{52} & \ding{52} &  \ding{52} & \ding{52} & \ding{52} & $\bullet$ & $\bullet$ & $\bullet$ \\
      $\Delta \eta (\mu,\tau_{h})$ & \ding{52} & \ding{52} & \ding{52} & \ding{52} & \ding{52} & \ding{52} & \ding{52} & \ding{52} & \ding{52}  & $\bullet$ & $\bullet$ \\ \hline 
    \end{tabular}
    \caption{Kinematic observables used in the BDT analysis for the $\mu\tau_{h}$ channel with different BSM Higgs boson masses. The used observables are marked with a \ding{52}~(shows the six observables with the highest rank) or a $\bullet$.}
    \label{tab:obs_rank_mutauj}
\end{table}

\begin{table}[!t]
{\scriptsize
    \centering\scalebox{0.99}{
    \begin{tabular}{|c|c|c|c|c|c|c||c|c|} \hline
         & \multicolumn{3}{c|}{$\mu\tau_{h}$} & \multicolumn{3}{c||}{$\mu\tau_{e}$} & \multicolumn{2}{c|}{$\mu\tau_{e} + \mu\tau_{h}$ } \\ \cline{2-9}
         & Signal &  & U.L. & Signal &  & U.L. & \multicolumn{2}{c|}{$2\sigma$ U.L. (fb)}\\ \cline{8-9}
         $m_{H}$ & eff. & Bkg. & ($2\sigma$) & eff. & Bkg. & ($2\sigma$) & \multicolumn{2}{c|}{$\kappa$} \\ \cline{8-9}
        (GeV) & ($10^{-3}$) &  & (in fb) & ($10^{-3}$) &  & (in fb) & 0 & $5\%$  \\  \hline 
        20 & 0.46 & 11 & 4.8 & 6.7 & 557 & 2.3 & 1.6 & 2.1 \\ \hline 
        40 & 4.4 & 26 &  0.77 & 5.4 & 250 & 1.9 & 0.55 & 0.60 \\ \hline 
        60 & 5.3 & 81 &  1.1 & 3.8 & 641 & 4.4 & 0.90 & 1.1 \\ \hline 
        80 & 2.1 & 32 &  1.8 & 5.6 & 1928  & 5.2 & 1.3 & 1.6  \\ \hline 
        100 & 3.5 & 67 &  1.6 & 7.1 & 463 & 2.0 & 0.88 & 1.1 \\ \hline 
        125 & 19.9 & 1603 &  1.3 & 9.4 & 1479 & 2.7 & 0.90 & 2.0\\ \hline 
        150 & 11.9 & 137 &  0.65 & 5.8 & 176 & 1.5 & 0.46 & 0.54 \\ \hline 
        200 & 25.5 & 235 &  0.40 & 7.8 & 115 & 0.92 & 0.28 & 0.34 \\ \hline 
        400 & 43.5 & 60 &  0.12 & 22.8 & 833 & 0.84 & 0.10 & 0.12 \\ \hline 
        600 & 79.8 & 55 &  $0.062$ & 26.3 & 235 & 0.39 & $0.053$ & $0.058$ \\ \hline 
        800 & 78.5 & 55 & $0.063$ & 30.6 & 170  & 0.28 & $0.052$ & $0.056$ \\ \hline 
    \end{tabular}}
    \caption{Signal efficiency, background yields, and projected upper limits on $\sigma(pp \to Hjj)\times {\rm Br}(H \to \mu\tau)$ at $2\sigma$, from searches for VBF Higgs production in the $\mu\tau_{h}$ and $\mu\tau_{e}$ channels at $\sqrt{s}=14~$TeV LHC with $\mathcal{L}=3~\rm{ab}^{-1}$.}
    \label{tab:cut_flow_bsm}
    }
\end{table}

\begin{figure*}[!t]
    \centering
    \includegraphics[width=0.495\textwidth]{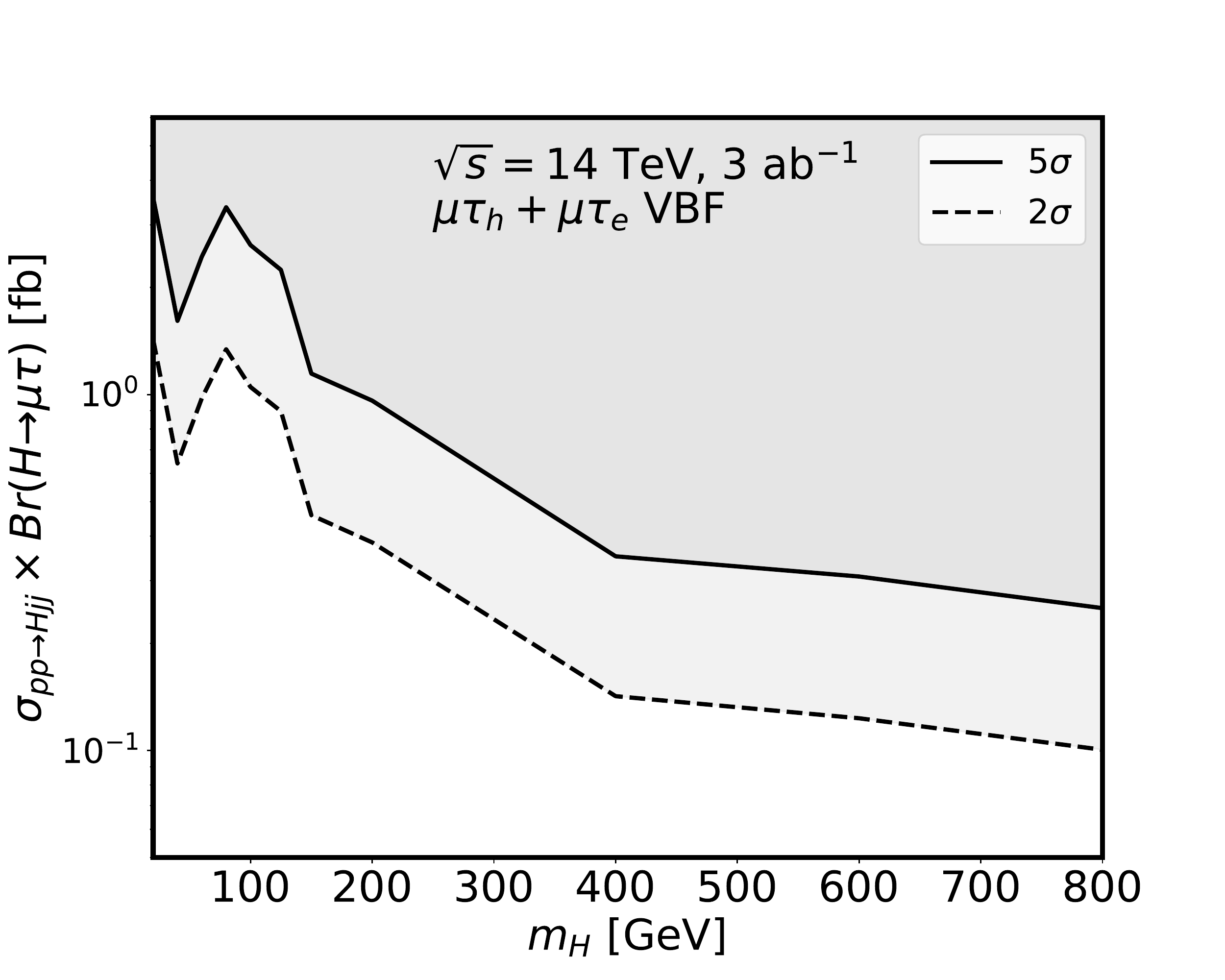}
    \includegraphics[width=0.495\textwidth]{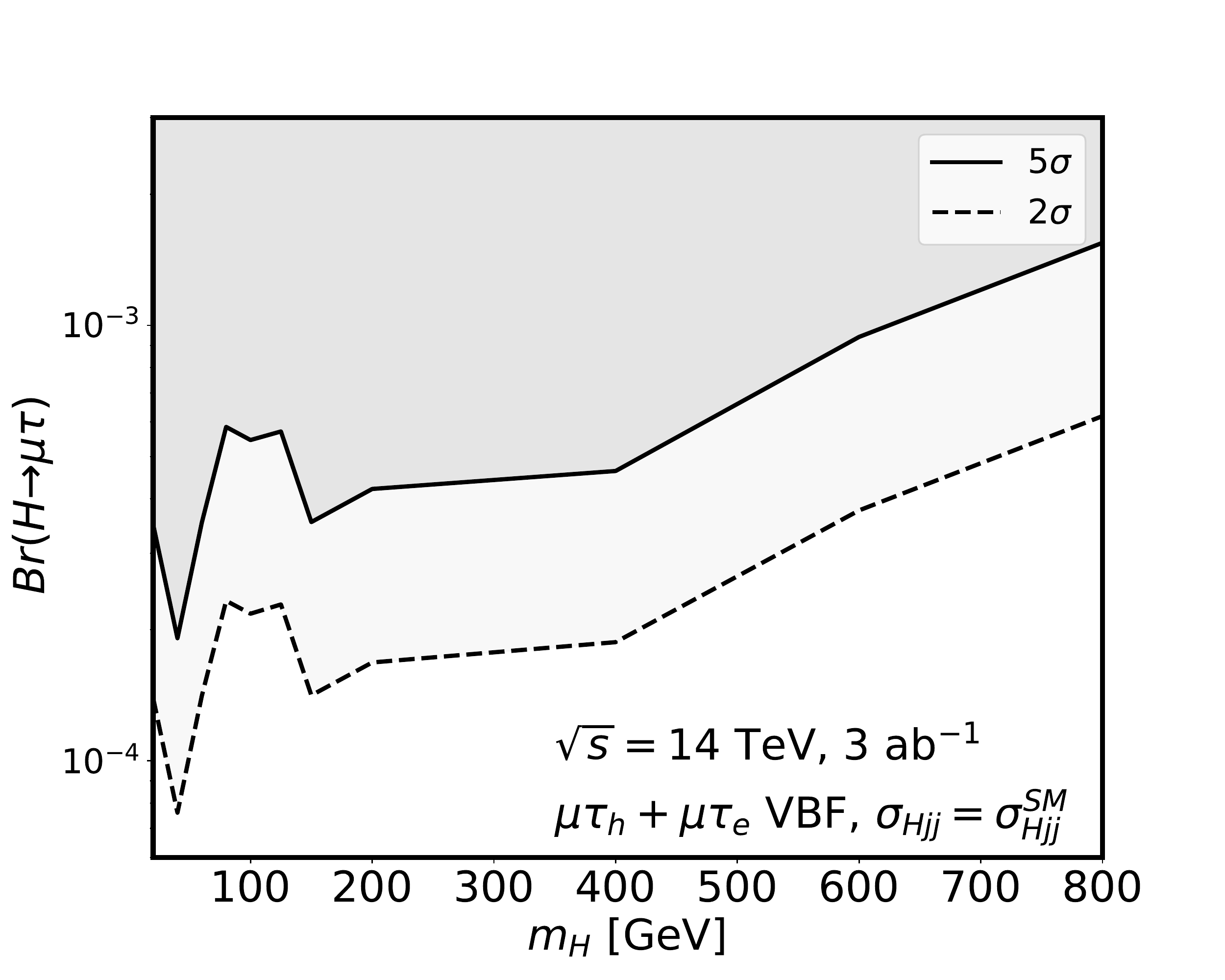}
    \caption{\textit{Left}: Projected upper limits on the production cross-section for the Higgs boson in the vector boson fusion mode times the branching ratio for LFV decays of the BSM Higgs boson, $\sigma_{H \to \mu\tau}^{\rm VBF}$, from combined searches in the $\mu\tau_{e} + \mu\tau_{h}$ channels at the $\sqrt{s}=14~$TeV at $\mathcal{L}=3~\rm{ab}^{-1}$. \textit{Right}: Projected upper limits on the left panel are translated into upper bounds on ${\rm Br}(H \to \mu\tau)$ considering SM-like vector boson fusion Higgs production rates. 
    }
    \label{fig:cs_br_ul}
\end{figure*}

The signal efficiency and background yields at the HL-LHC from the BDT analysis in the $\mu\tau_{h}$ and $\mu\tau_{e}$ channels for various signal benchmarks are presented in Table~\ref{tab:cut_flow_bsm}. We translate the results into projected upper limits on the VBF Higgs production cross-section times the branching ratio for LFV Higgs decays, $\sigma_{H \to \mu\tau}^{\rm VBF} = n_{S}\sqrt{B+(\kappa\cdot B)^{2}}/(\mathcal{L}\times \epsilon_{S}^{\rm VBF})$, as a function of $m_{H}$. Projected upper limits at $2\sigma$ from searches in the $\mu\tau_{e}$ and $\mu\tau_{h}$ channels are shown in Table~\ref{tab:cut_flow_bsm}. Limits from combined searches in $\mu\tau_{e}+\mu\tau_{h}$ channels are also shown for $\kappa=0$ and $5\%$. We observe that the projected limits get stronger with $m_{H}$ except around $m_{H} \sim m_{Z}$ and $m_{H} \sim m_{h_{\rm SM}}$. For $m_{H} \sim m_{Z}$, the signal has a greater overlap with the $(Z \to \tau\tau) + {\rm jets}$ background than the nearby signal benchmark points, which leads to a relatively lower signal efficiency. A similar albeit smaller decline in sensitivity occurs when the mass of the BSM Higgs boson is closer to $m_{h_{\rm SM}}$ due to greater overlap between the signal and the ggF/VBF $h \to \tau\tau$ backgrounds in the input dataset used for the BDT analysis, despite their sub-dominant contributions to the final event rate. 

The combined search limits are interpolated to derive projected upper bounds on $\sigma_{H \to \mu\tau}^{\rm VBF}$ for $m_{H} \in [20,800]~$GeV in Fig.~\ref{fig:cs_br_ul}~(left). It must be noted that the aforesaid projections are independent of any model considerations and can be translated to any new physics scenario in order to test their projected reach at the HL-LHC. We observe that the HL-LHC would be able to probe LFV Higgs decays in VBF Higgs production up to $\sigma_{H \to \mu\tau}^{\rm VBF} \gtrsim 0.28~(0.05)~$fb for $m_{H} = 200~(800)~$GeV, at $2\sigma$ uncertainty. The projected sensitivity drops down to 0.34 and 0.056 upon considering $5\%$ systematic uncertainty. Note that these projected limits are roughly two orders of magnitude smaller than the existing LHC limits on $\sigma(gg\to H)\times {\rm Br}(H\to \mu\tau)$~\cite{CMS:2019pex}.  

Considering SM-like VBF Higgs production rates, the above projections for $\kappa=0$ are translated into upper limits on ${\rm Br}(H \to \mu\tau)$ in Fig.~\ref{fig:cs_br_ul}~(right). The projected sensitivity for ${\rm Br}(H \to \mu\tau)$ at the HL-LHC is roughly $\sim 0.0015\%$ at $m_{H}=300~$GeV, which weakens to $\sim 0.0030\%$ as $m_{H}$ increases to $800~$GeV. At this point, we would like to note that the inclusion of ggF contributions to the VBF signal could further enhance the projected sensitivity, with the overall improvement being largely governed by the ratio of $\sigma_{H}^{\rm ggF}/\sigma_{H}^{\rm VBF}$.

\begin{figure}[!t]
    \centering
    \includegraphics[width=0.495\textwidth]{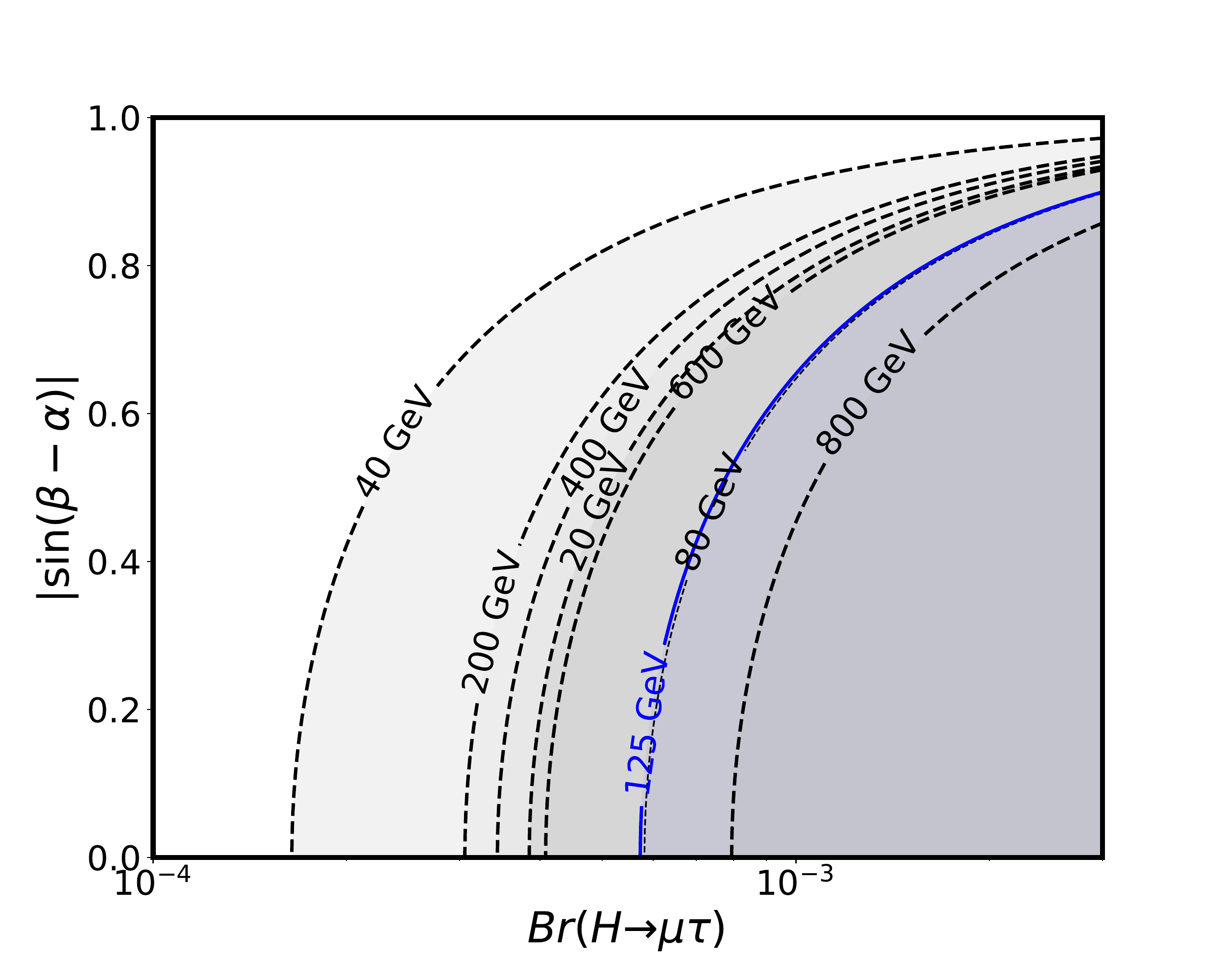}
    \caption{Projected reach at $5\sigma$ in the plane of $|\sin(\beta - \alpha)|$ and ${\rm Br}(H \to \mu\tau)$ for a generic 2HDM scenario for different BSM Higgs masses from combined searches in the $\mu\tau_{e} + \mu\tau_{h}$ channels with VBF Higgs production at the $\sqrt{s}=14~$TeV at $\mathcal{L}=3~\rm{ab}^{-1}$. The limit vanishes for $\sin(\beta-\alpha)\to 1$ (or $\cos(\beta-\alpha)\to 0$, the so-called alignment limit), in which case, the $HWW$ coupling vanishes and there is no VBF production of $H$.}
    \label{fig:2HDM_cos_br}
\end{figure}

To illustrate these constraints in a concrete model framework, we reinterpret them in a generic 2HDM~\cite{Lee:1973iz,Branco:2011iw} scenario, where the coupling of the SM-like Higgs boson $h$ and the BSM Higgs boson $H$ to weak bosons is $\sin(\beta - \alpha)$ and $\cos(\beta - \alpha)$ times the SM coupling, respectively, where $\beta$ is the ratio of vacuum expectation values for the two Higgs doublets and $\alpha$ is the neutral Higgs mixing angle. Accordingly, in the 2HDM scenario, the VBF BSM Higgs production cross-section can be parameterized as $\cos^{2}(\beta - \alpha)$ times the SM production rate. Using this analogy, we recast the upper limits on the VBF production cross section $\sigma(pp \to Hjj)$ times the branching ratio ${\rm Br}(H \to \mu\tau)$ into projections in the plane of $|\sin(\beta - \alpha)|$ and ${\rm Br}(H \to \mu\tau)$ for several BSM Higgs masses, as shown in Fig.~\ref{fig:2HDM_cos_br}. The shaded areas represent the projected exclusion regions at $5\sigma$ uncertainty for the respective values of $m_{H}$. It is important to note that the results in Fig.~\ref{fig:2HDM_cos_br} are directly correlated to the projections for ${\rm Br}(H \to \mu\tau)$ shown in Fig.~\ref{fig:cs_br_ul}~(right). We observe that among the different representative signal benchmarks considered in the present analysis, the strongest sensitivity for ${\rm Br}(H \to \mu\tau)$ and henceforth, in the $\{|\sin(\beta-\alpha), {\rm Br}(H \to \mu\tau)|\}$ plane for a generic 2HDM model is observed for $m_{H} = 40~$GeV. The sensitivity weakens as $m_{H}$ approaches $m_{Z}$ or $m_{h}$ (as shown by the blue curve) due to a larger overlap with the backgrounds. Interestingly, the area of the projected exclusion region in the plane of $\{|\sin(\beta-\alpha), {\rm Br}(H \to \mu\tau)|\}$ improves with increasing $m_{H}$ until around $m_{H}=150~$GeV beyond which the sensitivity again recedes gradually due to smaller cross sections.  

\begin{figure*}[!t]
    \centering
    \includegraphics[width=0.48\textwidth]{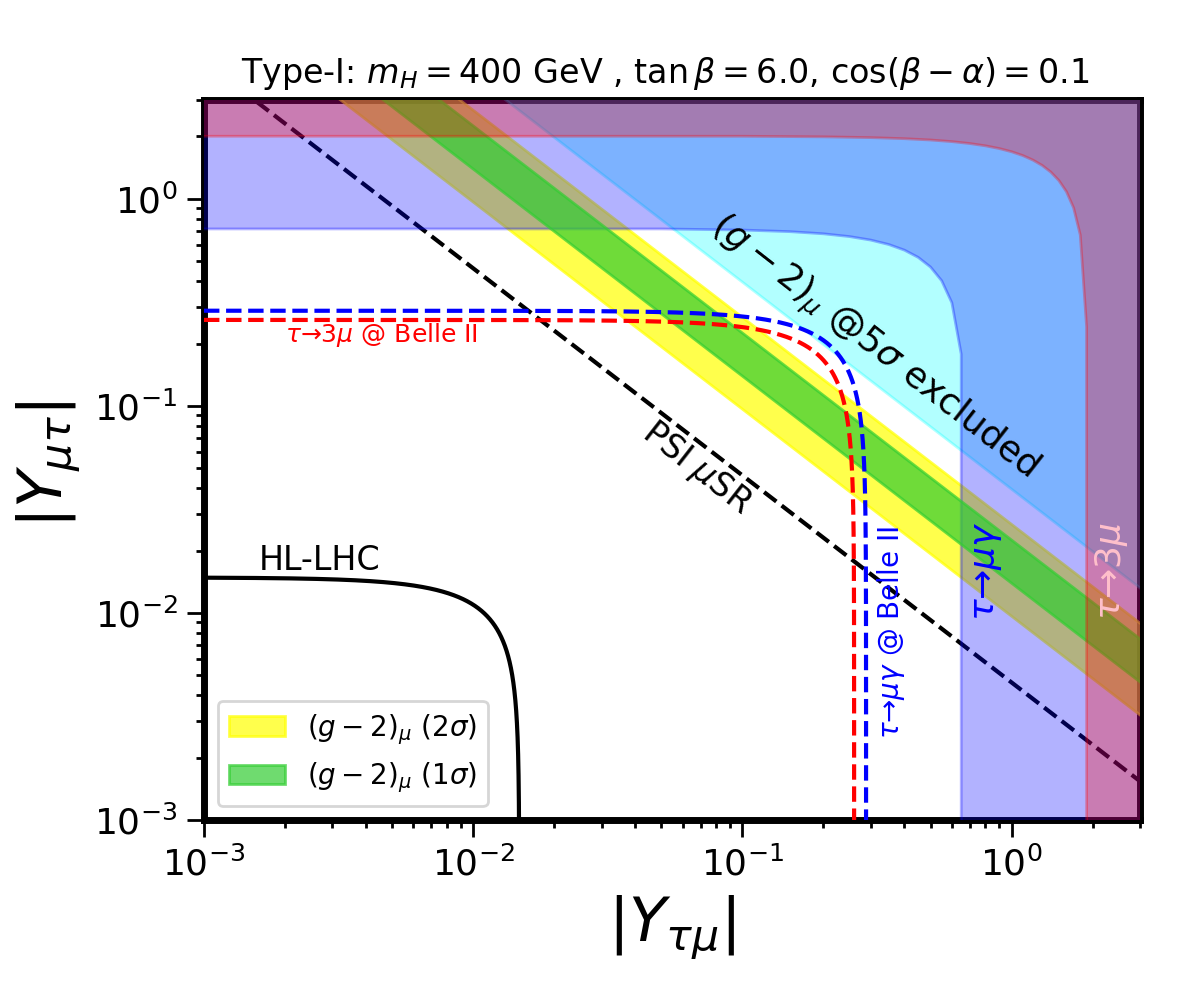}
    \includegraphics[width=0.48\textwidth]{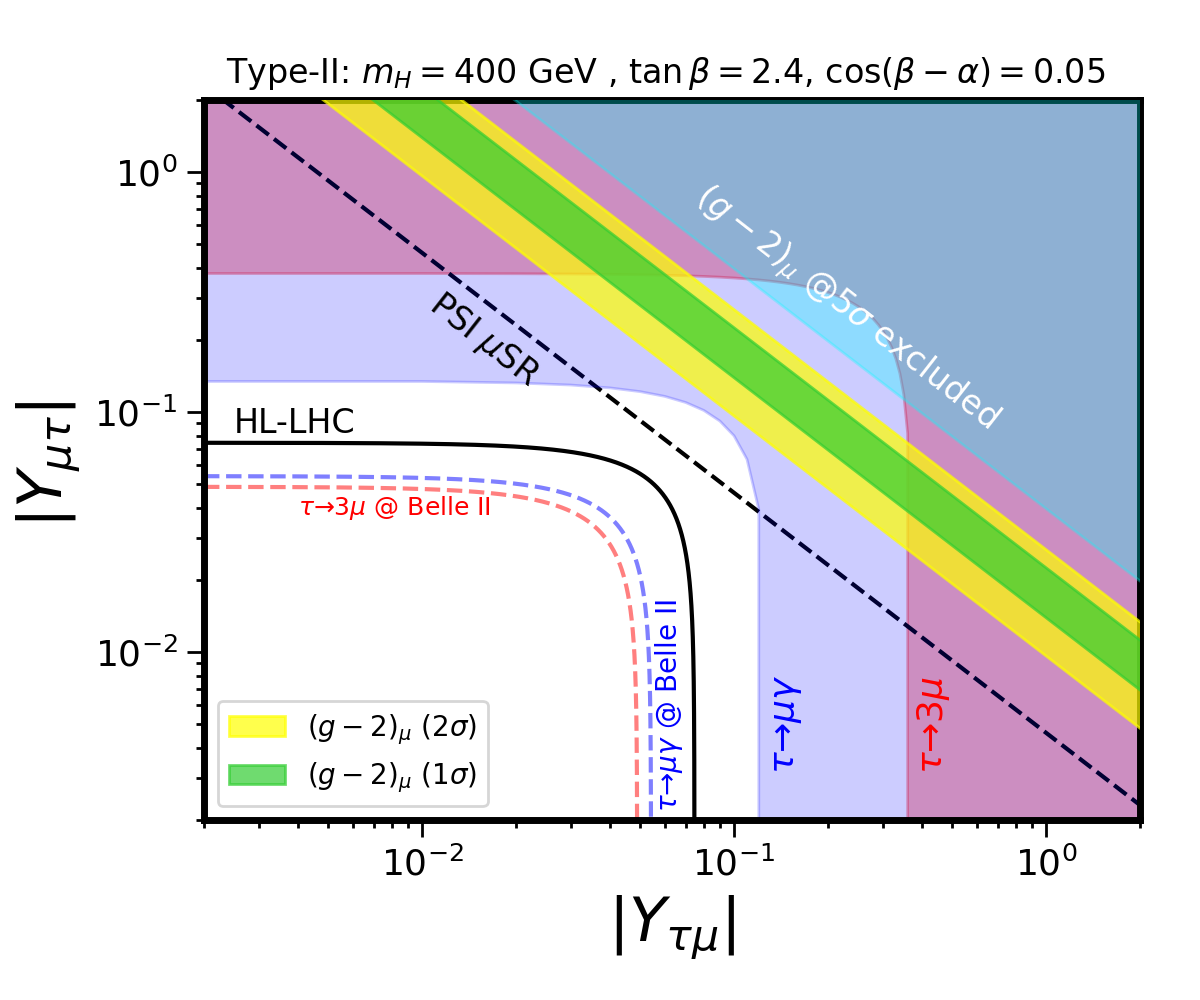}
    \caption{Projected upper limits on the off-diagonal Yukawa couplings $|Y_{\mu\tau}|$ and $|Y_{\tau\mu}|$ from direct searches in the combined $\mu\tau_e+\mu\tau_{h}$ channel at the HL-LHC are represented as solid black lines. The red (blue) shaded region represents the exclusion from LFV decays $\tau\to 3 \mu$~\cite{Hayasaka:2010np}  ($\tau \to \mu \gamma$~\cite{BaBar:2009hkt}), whereas the red and blue dashed lines are the corresponding future projected sensitivities at Belle II~\cite{Belle-II:2022cgf}. The green (yellow) band represents the preferred parameter space to accommodate the $(g-2)_\mu$ anomaly~\cite{Muong-2:2021ojo} at $1\sigma$ ($2\sigma$). The black dashed line represents the future sensitivity reach of $\mu$EDM~\cite{Adelmann:2021udj} for $\Im(Y_{\mu\tau}Y_{\tau\mu})$. }
    \label{fig:LFV_2HDM}
\end{figure*}


The above result can be easily reinterpreted in different avatars of the 2HDM with LFV Higgs bosons~\cite{Bjorken:1977vt, Davidson:2010xv, Branco:2011iw, Kopp:2014rva, Omura:2015nja, Botella:2015hoa, Buschmann:2016uzg, Altmannshofer:2016zrn, Primulando:2016eod, Crivellin:2019dun, Iguro:2019sly, Vicente:2019ykr, Primulando:2019ydt,  Ghosh:2020tfq, Ghosh:2021jeg, Blanke:2022kpi}. For illustration, we consider two representative benchmarks, BP$_I$ with $\cos(\beta - \alpha) = 0.1$, $\tan\beta = 6$, $m_{H} = 400~$GeV in the Type-I 2HDM, and BP$_{II}$ with $\cos(\beta - \alpha) = 0.05$, $\tan\beta = 2.4$, $m_{H} = 400~$GeV in the Type-II 2HDM scenario,  respectively. Both benchmarks are chosen from the parameter space allowed by the current Higgs global fit constraints~\cite{Kling:2020hmi}. We study the projected sensitivity for BP$_I$ and BP$_{II}$ from searches in the $\mu\tau_{e} +\mu\tau_{h}$ channels with VBF heavy Higgs production at the HL-LHC, using the results from our collider analysis~(see Fig.~\ref{fig:cs_br_ul}). The projected sensitivities at the HL-LHC are presented in the plane of $|Y_{\mu\tau}|$ and $|Y_{\tau\mu}|$ in Fig.~\ref{fig:LFV_2HDM} as the solid-black curves. These are analogous to but weaker than the numbers shown in  Fig.~\ref{fig:LFV_Sm_Higgs} for the 125 GeV Higgs case, mainly because the VBF production cross section for the Heavy Higgs boson is suppressed by a $\cos^2(\beta-\alpha)$ factor, as compared to the SM-like case. 

In Fig.~\ref{fig:LFV_2HDM}, the parameter space constrained by LFV decays $\tau\to 3\mu$~\cite{Hayasaka:2010np} and $\tau \to \mu\gamma$~\cite{BaBar:2009hkt} is shaded in red and blue, respectively, while their future reach at BELLE-II~\cite{Belle-II:2022cgf} is shown by the dashed curves. The green (yellow)-shaded area represents the parameter region consistent with the $(g -2)_{\mu}$ anomaly at $1(2)\sigma$. Here again, we observe that the projected sensitivities at the HL-LHC exceed the future LFV reach of BELLE-II measurements for BP$_I$ and they are comparable for BP$_{II}$. It is also worth noting that, unlike in the SM Higgs scenario, LFV can be further suppressed here by an appropriate choice of $\alpha$ and $\beta$, whereas the HL-LHC sensitivity is primarily determined by the combination $\beta-\alpha$. Moreover, for BP$_I$, we find that there is currently some allowed parameter space to explain the $(g -2)_{\mu}$ anomaly, which can be completely probed by HL-LHC.

Besides the 2HDM model, there are several other well-motivated extended Higgs sector models where LFV decays might naturally arise, such as in the little Higgs models~\cite{Yang:2016hrh,delAguila:2019htj,Pacheco:2021djh, Ramirez:2022zpk}, left-right symmetric models~\cite{Dev:2016dja,Maiezza:2016ybz,Dev:2016nfr,Dev:2017dui, Boyarkin:2018wlr}, mirror models~\cite{Hung:2006ap,Hung:2007ez,Bu:2008fx,Chang:2016ave,Hung:2017voe}, supersymmetric models~\cite{Diaz-Cruz:1999sns, Han:2000jz,Diaz-Cruz:2008agf,Arhrib:2012ax, Arana-Catania:2013xma,Aloni:2015wvn,Zhang:2015csm,Demidov:2016gmr,Dong:2018rfs, Zhang:2021nzv,Wang:2022tgf}, models with flavor symmetries~\cite{Ishimori:2010au}, composite Higgs models~\cite{Agashe:2009di,Azatov:2009na}, warped extra-dimension models~\cite{Perez:2008ee, Casagrande:2008hr, Blanke:2008zb, Albrecht:2009xr, Buras:2009ka},  models with higher-dimensional operators~\cite{Diaz-Cruz:1999sns,Babu:1999me, Giudice:2008uua, Aguilar-Saavedra:2009ygx,Goudelis:2011un}, neutrino mass models~\cite{Pilaftsis:1992st,Korner:1992zk,Arganda:2014dta,Arganda:2015naa,Thuc:2016qva,Aoki:2016wyl,Arganda:2016zvc,Herrero-Garcia:2017xdu, Thao:2017qtn,Enomoto:2019mzl,Babu:2019mfe,Marcano:2019rmk,Haba:2020lqv,Babu:2020hun,Barman:2021xeq,Hundi:2022iva, Julio:2022bue,Dcruz:2022dao} and other models~\cite{Dorsner:2015mja,Baek:2015fma,Hue:2015fbb,Alvarado:2016par,Evans:2019xer,Hong:2020qxc}. The analysis presented in this section can be extended to these scenarios as well. 

\section{Conclusion}\label{sec:conclusion}

In this paper, we have examined the parameter space for LFV couplings of a neutral Higgs boson to muon and tau leptons. We show that these flavor violating couplings can be effectively probed at the HL-LHC through the VBF production process: $pp \to h j j \to (h \to \mu \tau) jj$, complementary to the ggF process. For the SM Higgs, we find $\sqrt{|Y_{\mu\tau}|^{2} + |Y_{\tau\mu}|^{2}} < 5.1\times 10^{-4}$ as the projected limit on the LFV Higgs Yukawa couplings, which turns out to be stronger than the existing and future limits from low-energy LFV observables like $\tau\to \mu\gamma$ and $\tau\to 3\mu$ (see Fig.~\ref{fig:LFV_Sm_Higgs}).  In addition, we have also studied the LFV arising from a generic BSM neutral Higgs boson $H$ in the mass range of $m_{H}\in [20,800]~$GeV and have given the projected model-independent upper limits on the VBF production cross-section of $Hjj$ times the branching ratio of $H\to \mu\tau$ at the HL-LHC (see Fig.~\ref{fig:cs_br_ul}). Finally, we have reinterpreted our results for the BSM neutral CP-even Higgs boson in the 2HDM (see Figs.~\ref{fig:2HDM_cos_br} and \ref{fig:LFV_2HDM}). We find that HL-LHC will provide comparable or better constraints than the low-energy LFV searches on the LFV couplings of the BSM Higgs boson.

\section*{Acknowledgments} 
RKB thanks Biplob Bhattacherjee and Terrance Figy for helpful discussions. The work of RKB is supported by the U.S. Department of Energy under Grant No. DE-SC0016013. Some of the computing for this project was performed at the High Performance Computing Center at Oklahoma State University, supported in part through the National Science Foundation Grant No. OAC-1531128. The work of BD was supported in part by the U.S. Department of Energy under Grant No.~DE-SC0017987 and by a URA VSP Fellowship. BD acknowledges the Fermilab theory group for local hospitality during the completion of this work. AT acknowledges the Aspen Center of Physics for local hospitality during the completion of this work. The work  at the Aspen Center for Physics is supported by National Science Foundation Grant No. PHY-1607611. 
\bibliographystyle{utcaps_mod}
\bibliography{references}

\end{document}